\documentclass[mathematics,article,accept,moreauthors,pdftex,12pt,a4paper]{mdpi}
\lastpage{x}
\doinum{10.3390/------}
\pubvolume{xx}
\pubyear{2014}
\history{Received: 12 April 2014; in revised form: 24 May 2014 / Accepted: 15 August 2014 / \\Published: xx}

\usepackage{perpage}

\usepackage{soul}
\usepackage{amssymb,amsmath}
\usepackage{graphicx}
\usepackage{subfigure}
\makeatletter
\renewcommand{\@thesubfigure}{\normalsize(\textbf{\alph{subfigure}})}
\makeatother
\usepackage{array, dcolumn, multirow}
\usepackage{hyphenat}
\usepackage{enumitem,caption,booktabs}
\setitemize{parsep=0pt,itemsep=0pt,leftmargin=.4in}
\setenumerate{parsep=0pt,itemsep=0pt,leftmargin=.4in}

\MakePerPage{footnote} 
\usepackage{url}
\usepackage{amsfonts}%

\usepackage{color}
\usepackage{epstopdf}

\renewcommand{\thefootnote}{\fnsymbol{footnote}}

\Title{Modeling the Influence of Environment and Intervention on Cholera in Haiti}

\Author{Stephen Tennenbaum *, Caroline Freitag and Svetlana Roudenko }

\address[1]{Department of Mathematics, The George Washington University, 2115 G St. NW, Washington, \linebreak DC 20052, USA;  E-Mails:  calliefreitag@gmail.com (C.F.); roudenko@gwu.edu (S.R.)}

\corres{E-Mail: set1@gwu.edu; \linebreak Tel.: +1-202-994-6235.}

\abstract{We propose a simple model with two infective classes in order to model the
cholera epidemic in Haiti. We include the impact of environmental events
(rainfall, temperature and tidal range) on the epidemic in the Artibonite and
Ouest regions by introducing terms in the transmission rate that vary with
environmental conditions. We fit the model on weekly data from the beginning
of the epidemic until December 2013, including the vaccination programs that were
recently undertaken in the Ouest and Artibonite regions. We then modified
these projections excluding vaccination to assess the programs' effectiveness.
Using real-time daily rainfall, we found lag times between precipitation events
and new cases that range from $3.4$ to $8.4$ weeks in Artibonite and $5.1$ to
$7.4$ in Ouest. In addition, it appears that, in the Ouest region, tidal
influences play a significant role in the dynamics of the disease. Intervention efforts of all types have reduced case numbers in both regions; however, persistent outbreaks continue. In Ouest, where the population at risk seems particularly besieged and the overall population is larger, vaccination
efforts seem to be taking hold more slowly than in Artibonite, where a smaller
core population was vaccinated. The models including the vaccination
programs predicted that a year and six months later, the mean number of cases
in Artibonite would be reduced by about two thousand cases, and in Ouest by twenty four hundred cases below that predicted by the models without vaccination. We also
found that vaccination is best when done in the early spring, and as early as
possible in the epidemic. Comparing vaccination between the first spring and
the second, there is a drop of about $40\%$ in the case reduction due to the
vaccine and about $10\%$ per year after that.
\vspace{24pt}
}


\begin{document}
\vspace{-12pt}


\section{Introduction}

On 12 January 12 2010, a $7.0$ magnitude earthquake struck near Haiti's capital,
Port-au-Prince \cite{WHO}. \linebreak The poorest nation in the Western Hemisphere, the
earthquake shattered Haiti's already weak infrastructure \cite{WHO}. Thousands
of Haitians were killed, and even more were forced to flee to resettlement
camps \cite{WHO}.

In October 2010, the first case of cholera, ever on record, was reported in Haiti. A later UN
investigation revealed the specific strain of \emph{V. cholerae} came from
South Asia \cite{Cravioto}. The UN investigation and epidemiological
literature suggest that the epidemic began outside of a UN peacekeeper camp
near Mirebalais in the Centre department, along the Artibonite River
\cite{Cravioto, Piarroux}. As \emph{V. cholerae} is a waterborne pathogen, the
Artibonite River is the ostensible route through which the disease spread
throughout Haiti's ten administrative regions, called departments \cite{MSPP}.

Anecdotal news reports describe the dismal situation for thousands of Haitians
that still remain displaced months after the earthquake. Sewage of millions of
people flow through open ditches. Human waste from septic pits and latrines is
dumped into the canals and, after it rains, ends up in the sea. Those living
close to the water use over-the-sea toilets, and next to these outhouses,
fishing boats unload and sell the fish from plastic buckets, \textit{etc}.
\cite{Knox2012b}.

Haiti's two most populous regions, Ouest and Artibonite, were also the two
regions hardest hit by the epidemic. Cases in Ouest and Artibonite account for
60\% of the total burden of cholera in Haiti \cite{PAHO}. For this reason, we
chose to focus our analysis on the Ouest and Artibonite regions. By 7 April
2012, cholera had affected $5.7\% $ of the total population  in Ouest and $6.9\% $ of the population in Artibonite \cite{PAHO}  (note:
population data for Haiti is from 2009 \cite{MSPP}, one year before the
earthquake).

\subsection{Previous Research}

Previous models dealing with cholera and climatic conditions (rainfall,
precipitation and tides) in Haiti vary widely in approach. There are a number
of dynamical models using variations of SIWR (Susceptible,
Infected, contaminated Water, Recovered),
proposed in 2001 by Code\c{c}o \cite{Codeco} and, later, Tien and \scalebox{.95}[1.0]{Earn
\cite{Tien2010}, that address the situation in Haiti \cite{Andrews, Bertuzzo,
Tuite}.} These models look at various compartmental and spatial structures, but
do not take environmental conditions explicitly into account. The
models proposed by Tuite, \emph{et al}. \cite{Tuite} and Bertuzzo, \emph{et al}.
\cite{Bertuzzo}, for example, both incorporated a \textquotedblleft
gravity\textquotedblright\ term to study the interaction among departments.
The model proposed by Andrews and Basu \cite{Andrews} accounted for a
bacterial \textquotedblleft hyperinfectivity\textquotedblright\ stage,
following research by Hartley, \emph{et al}. in 2006 showing that \linebreak \emph{V. cholerae}
initially has a higher infectivity before it decays to a lower infective rate
in the aquatic reservoir. A paper by Chao \emph{et al}. \cite{Chao} uses an agent-based model to investigate hypothetical vaccination programs in Haiti. This
model examined various vaccination strategies that included pre-vaccination
and early (21 days after the epidemic begins) reactive vaccination using
various strategies.

Other papers \cite{Reiner, Rinaldo, Eisenberg} do take precipitation directly
into account in cholera. The first \cite{Reiner} is a spatiotemporal Markov
chain model using seasonal rainfall that drives disease outbreaks in an urban
core, which then propagates to other areas of the city. The second
\cite{Rinaldo} deals specifically with Haiti and was done by the same group
that produced one of the earlier papers \cite{Bertuzzo}. In \cite{Rinaldo},
they looked at the reliability of the earlier studies, and they found that
although those models do well in capturing the early dynamics of the epidemic,
they fail to track latter recurrences forced by seasonal patterns
\cite{Rinaldo}. As a follow up, Rinaldo \emph{et al}. \cite{Rinaldo} add a
precipitation forcing function to their original model along with other
modifications, such as the river network and population mobility. These
modifications produce a better fit to the observed pattern of cases over the
first year of the epidemic. The third model by \linebreak Eisenberg \emph{et al}.
\cite{Eisenberg} looks at the link between precipitation and disease outbreaks
in Haiti from a statistical and dynamical modeling approach. Their dynamical
system is a hybrid SIWR-SIR approach, where the infection rate of the SIWR
component includes a rainfall forcing function and the infection rate of the
SIR component accounts for short-term direct contacts with infected individuals.

In addition, these models assessed the impact of potential intervention
strategies, including vaccination. Bertuzzo found that a vaccination campaign
aiming to vaccinate 150,000 people \linebreak after 1 January 2011, would have little
effect, in part because of the late timing and in part because of the large
proportion of asymptomatic individuals who would need to get vaccinated
\cite{Bertuzzo}. Both the models proposed by Tuite, \emph{et al}. \cite{Tuite} and
Andrews and Basu \cite{Andrews} suggest that vaccination campaigns would have
a modest effect. In March 2012, Partners in Health began vaccinating 100,000
individuals with Shanchol, a two-dose cholera vaccine \cite{Knox2012a}. The
size of the campaign was limited by the size of the global stockpile of
Shanchol \cite{Knox2012a}. The vaccination campaign is targeted at 50,000
individuals living in the slums of Port-au-Prince (Ouest region), where population density is
thought to increase the rate of cholera exposure, and at 50,000 individuals
living in the Artibonite River valley (Artibonite region), where the epidemic began
\cite{Knox2012a}. Chao \emph{et al}. \cite{Chao} showed that a targeted vaccination
strategy would have the best results for this limited supply of vaccine, and
by early vaccinating 30\% of the population and hygienic improvements, the
cases could be reduced by as much as 55\%.

In 2001, Code\c{c}o proposed introducing an oscillating term to model seasonal
variability \cite{Codeco}. However, none of the Haiti-specific models
published prior to 2012 account for seasonality. Haiti experienced flooding
in June 2011, October 2011, and March 2012 \cite{UN2012}. As cholera reached
an endemic state in Haiti, connecting precipitation explicitly to disease
dynamics became more suitable. Mathematical models incorporated seasonality in
order to more accurately predict the course of the epidemic and to simulate
the effects of potential interventions. In April 2012, Rinaldo \emph{et al}.
\cite{Rinaldo} reexamined the above four models (including their own
\cite{Bertuzzo}) and concluded that, among other factors, seasonal rainfall
patterns were necessary to account for resurgences in the epidemic. They use
long-term monthly averages to augment the bacterial growth term of
contaminated water-bodies. Eisenberg \emph{et al}. \cite{Eisenberg} examined rainfall
patterns and assessed lag times between precipitation events and cases in the
early epidemic. These indicated short delays of four to seven days. Other papers
dealing with environmental factors were: \linebreak (1) a study of cholera in Zanzibar,
East Africa that demonstrated an eight-week fixed delay \cite{Reyburn} between
rainfall and cholera outbreaks; and (2) a study in Bangladesh \cite{deMagny}
that reports a somewhat shorter delay (four weeks). Both of these studies use a
statistical approach with seasonal data. Both also made note of the potential
influence of ocean environmental factors, and the Reyburn \emph{et al}. paper
included sea surface height and sea surface temperature in their analysis, but
failed to find any significant \linebreak relationship \cite{Reyburn}. In a third paper
Koelle \emph{et al}. (2005) \cite{Koelle} model very long time periods, more than a
year, in Bangladesh. This model also uses seasonal precipitation and models
changes in the susceptible fraction of the population due to demographics and
loss of immunity.

\subsection{The Model}

In this paper, we use detailed and current rainfall, temperature, and
predicted tides to model cholera in the Artibonite and Ouest regions.
This paper is the first, that we know of, that uses tidal range in a
model of cholera dynamics. We forego a bacterial or contaminated water compartment in favor of a
saturating infectious compartment with a time delay. This has the advantage of
more tractable temporal estimates without over parameterizing and including
compartments that are essentially unmeasurable.

These long-term trends and environmental influences establish the pattern of
response of the epidemic in Artibonite and Ouest. Thus, parameters were chosen
and model calibration set prior to a vaccination program being implemented. We
then used the model to evaluate the performance of the vaccination program
against the backdrop of an alternative history without vaccination.


\section{Material and Methods}

For Ouest and Artibonite, we investigated the correlations between reported
cholera cases and rainfall, temperature and, in the case of Ouest, tidal
range. We wanted to determine.
\begin{itemize}{ \item{if such correlations exist}, \item{the time
delay between environmental conditions and recorded cholera outbreaks}, and \item{if the effectiveness of a recent vaccination program could be assessed by use of this model.}}\end{itemize}

\subsection{Model}

\begin{figure}[H]
\centering
\includegraphics[width=6in,height=4.4in]{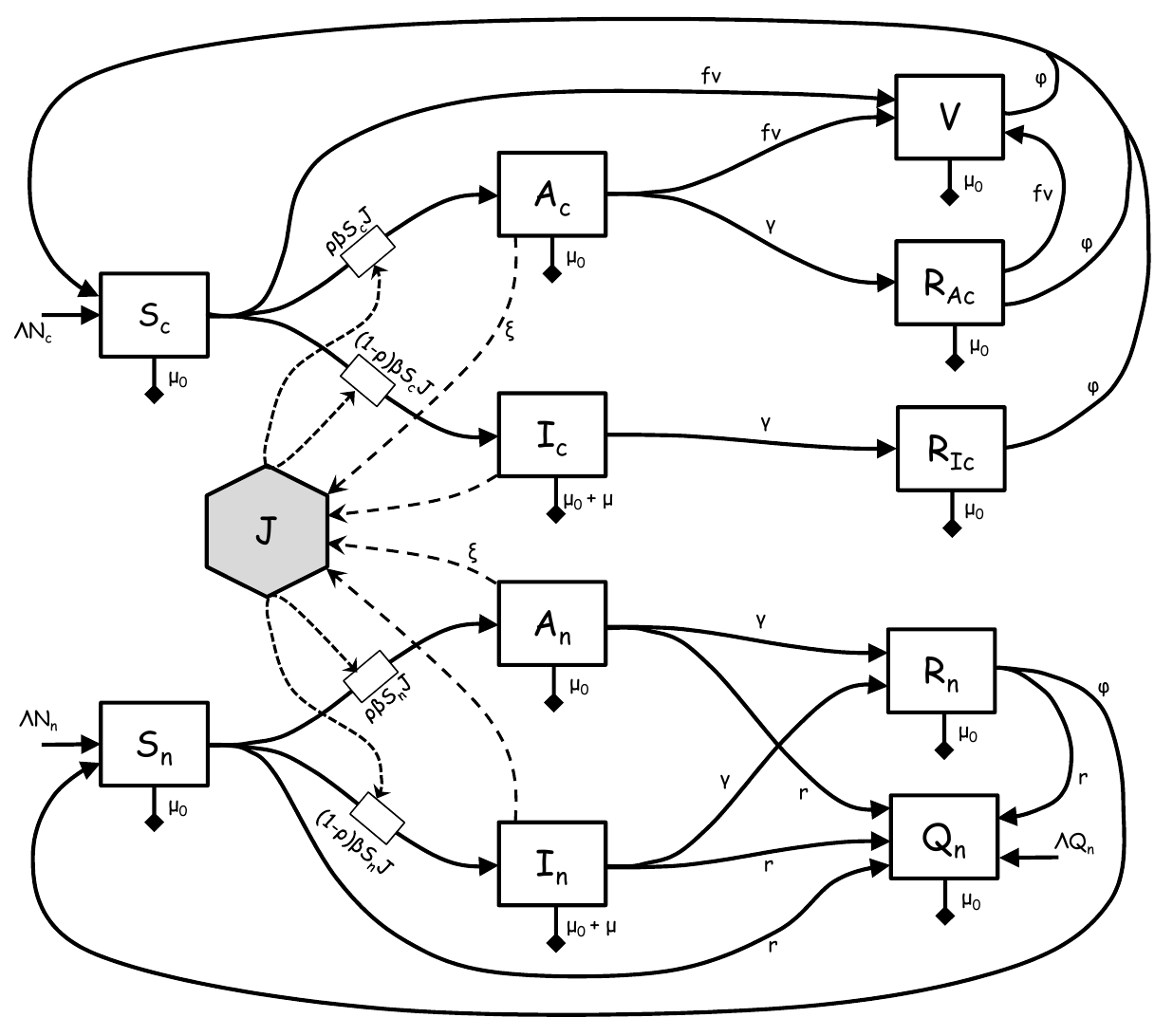}
\caption{A
compartmental model describing the movement of individuals from susceptible to
infectious to recovered, with two infectious classes, symptomatic and
asymptomatic; a transitional main population in which conditions
improve over time and an intractable indigent core. Compartment $J$ is the
time-delayed effective infectious class that replaces explicit water
bodies and bacterial compartments (see text).}%
\label{Fig0}%
\end{figure}

Our modeling approach is a modification of SIR type models, where individuals move
from susceptible $(S)$ to infected to recovered $(R)$
classes. Conceptually, the underlying model is a variation on the SIWR
model, which assumes that cholera is spread through susceptibles' contact with
contaminated water, food or fomites. SIWR models use the amount of water
consumed as a proxy for all possible modes of transmission, and the
concentration of bacteria in the water consumed modifies the infection rate by
a dose-response expression (see \cite{Tien2010} with a base model
\cite{Codeco}). An infectious individual may either be symptomatic
$(I)$ or asymptomatic $(A)$ \cite{Lenhart}. The probability, $\rho,$ of
asymptomatic infection is \linebreak $0.79$ \cite{Lenhart, Hartley}. Both
symptomatic and asymptomatic individuals move to the recovered
group, $R$, at a rate $\gamma$. Symptomatic individuals die from
cholera at a rate $\mu$. Transmission rates, $\beta\!\left( t\right) $, are estimated by fitting the number of cases predicted by the model (both incidence and cumulative cases) to the data. A schematic of the system is given in Figure \ref{Fig0}.

We chose not to incorporate water bodies and environmental bacterial populations explicitly, since this requires the estimation of extra compartments and a half dozen or so other parameters for which there are no data. The overall impact of having water bodies and bacteria concentrations is that there is a saturating effect on the force of infection, since the probability of getting an infectious dose from contact with a given body of water (or other sources) necessarily asymptotes as the concentration of bacteria increases (as an absolute limit, it cannot exceed one). Thus, it seems reasonable to replace water bodies and bacteria with a saturating function $\left( J \right) $ of the time delayed number of infected people; this has the advantage of eliminating superfluous variables and parameters, while retaining the essential dynamics. Time lags are also included in the action of precipitation and tides. These and all other parameters are discussed below, and values are given in Tables \ref{Table 1} and \ref{Table 2}.

\begin{table}[H]
\centering
\small

\begin{tabular}
[c]{ccccc}\toprule

\textbf{Parameter} & \textbf{Artibonite} & \textbf{Ouest} & \textbf{Units or Calculation} &
\textbf{References}\\\midrule

$\rho$ & $0.79$ & $0.79$ & fraction becoming asymptomatic & \cite{Lenhart,
Kaper}\\
$\gamma$ & $1.4$ & $1.4$ & fraction recovered per week & \cite{Andrews,
Bertuzzo, Codeco}\\
$\varphi$ & $0.01$ & $0.01$ & fraction losing immunity per week & \cite{Fung,
Legros}\\
$N_{0}$ & $1,571,020$ & $3,664,620$ & $-$ & \cite{MSPP}\\
$\Lambda$ & $0.00045$ & $0.00045$ & fraction births per week & \cite{CIA}%
\\
$\mu_{0}$ & $0.00015$ & $0.00015$ & fraction deaths (non-cholera) per week &
\cite{CIA}\\
$S_{0}$ & $1,534,338$ & $3,663,699$ & $N-\left( A_{0}+I_{0}+R_{0}%
+D_{0}\right) $ & \\
$I_{0}$ & $7653$ & $193$ & regression on early hospitalized cases &
\cite{PAHO}\\
$A_{0}$ & $28,790$ & $726$ & $\frac{\rho}{1-\rho}I_{0}$ & \\
$R_{0}$ & $0$ & $0$ & $-$ & \\
$D_{0}$ & $239$ & $2$ & $-$ & \cite{PAHO}\\
$t_{0}$ & 17 October 2010 & 17 October 2010 & 290${\text{th}}$ day of the year &
\cite{PAHO}\\\bottomrule

\end{tabular}

\caption{ Parameter values obtained from the literature, simple calculation or by
definition (sources are in the last column).}%
\label{Table 1}%
\end{table}


\begin{table}[H]%
\centering
\small
\begin{tabular}
[c]{ccccc}\toprule

\textbf{Parameter} & \textbf{Artibonite} & \textbf{Ouest} & \textbf{Units} & \textbf{Description}\\\midrule

$u_{0}$ & $0.091\;\left( 0.02\right) $ & $0.144\;\left( 0.02\right) $ &
unitless & \multicolumn{1}{l}{initial core fraction.}\\
$\theta_{p}$ & $4.980\;\left( 0.17\right) $ & $2.290\;\left( 0.27\right)
$ & weeks & \multicolumn{1}{l}{averaging window for precipitation}\\
$\tau_{p}$ & $3.381\;\left( 0.05\right) $ & $5.140\;\left( 0.21\right) $
& weeks & \multicolumn{1}{l}{delay for precipitation effects}\\
$k$ & $0.441\;\left( 0.13\right) $ & $0.373\;\left( 0.13\right) $ &
unitless & \multicolumn{1}{l}{temperature-precipitation interaction level}\\
$\alpha$ & $9.544\;\left( 1.45\right) \times10^{-3}$ & $2.200\;\left(
0.62\right) \times10^{-3}$ & week$^{-\text{1}}$ &
\multicolumn{1}{l}{infection rate}\\
$\theta_{m}$ & $-$ & $0.984\;\left( 0.69\right) $ & weeks &
\multicolumn{1}{l}{averaging window for tidal range}\\
$\tau_{m}$ & $-$ & $1.960\;\left( 0.47\right) $ & weeks &
\multicolumn{1}{l}{delay for tidal range effects}\\
$c$ & $-$ & $0.147\;\left( 0.28\right) $ & unitless &
\multicolumn{1}{l}{effect of tide relative to rain}\\
$M_{0}$ & $-$ & $25.74\;\left( 13.5\right) $ & cm &
\multicolumn{1}{l}{baseline of tidal range effect}\\
$r$ & $0.0562\;\left( 0.006\right) $ & $0.0376\;\left( 0.005\right) $ &
week$^{-\text{1}}$ & \multicolumn{1}{l}{decrease in non-core per
week}\\
$K_{I}$ & $\sim$$0$ & $291.69$ $\left( 117.9\right) $ & number &
\multicolumn{1}{l}{half saturation constant}\\
$\nu$ & $0.675$ $\left( 1.03\right) $ & $0.703$ $\left( 1.95\right) $ &
unitless & \multicolumn{1}{l}{efficacy of vaccine}\\\bottomrule

\end{tabular}

\caption{{{.
Numbers in parentheses are the 95$\%$ confidence intervals (CIs).}}}%
\label{Table 2}%
\end{table}

This paper evaluates Artibonite and Ouest separately, in order to capture the
different dynamics in each region. In each case, we use the following system of
non-autonomous ordinary differential equations and the components to be
fitted: $t$ is time since the beginning of the epidemic (week starting
17 October 2010) and $\Delta t$ is the time interval for reporting new cases
(one week).$\medskip$
\[%
\begin{array}
[c]{c}%
\text{\textbf{System of equations.
}}\\
\text{Core population.}\medskip\\
\left\{
\begin{array}
[c]{l}%
\frac{dS_{c}}{dt}=\Lambda N_{c}-\beta\!\left( t\right) S_{c}J-f\nu\!\left(
t\right) \frac{S_{c}}{S_{c}+A_{c}+R_{Ac}}-\mu_{0}S_{c}+\varphi\left(
R_{Ac}+R_{Ic}+V\right) \medskip\\
\frac{dA_{c}}{dt}=\rho\beta\!\left( t\right) S_{c}J-f\nu\!\left( t\right)
\frac{A_{c}}{S_{c}+A_{c}+R_{Ac}}-\left( \gamma+\mu_{0}\right) A_{c}%
\medskip\\
\frac{dI_{c}}{dt}=(1-\rho)\beta\!\left( t\right) S_{c}J-(\gamma+\mu+\mu
_{0})I_{c}\medskip\\
\frac{dR_{Ac}}{dt}=\gamma A_{c}-\left( \varphi+\mu_{0}\right) R_{Ac}%
-f\nu\!\left( t\right) \frac{R_{Ac}}{S_{c}+A_{c}+R_{Ac}}\medskip\\
\frac{dR_{Ic}}{dt}=\gamma I_{c}-\left( \varphi+\mu_{0}\right) R_{Ic}%
\medskip\\
\frac{dV}{dt}=f\nu\!\left( t\right) -\left( \varphi+\mu_{0}\right)
V\medskip\\
N_{c}=S_{c}+A_{c}+R_{Ac}+R_{Ic}+V\medskip
\end{array}
\right\}
\end{array}
\]%
\[%
\begin{array}
[c]{c}%
\text{Non-core population.}\medskip\\
\left\{
\begin{array}
[c]{l}%
\frac{dS_{n}}{dt}=\Lambda N_{n}-\beta\!\left( t\right) S_{n}J-rS_{n}-\mu
_{0}S_{n}+\varphi R_{n}\medskip\\
\frac{dA_{n}}{dt}=\rho\beta\!\left( t\right) S_{n}J-\left( \gamma+\mu
_{0}+r\right) A_{n}\medskip\\
\frac{dI_{n}}{dt}=(1-\rho)\beta\!\left( t\right) S_{n}J-(\gamma+\mu+\mu
_{0}+r)I_{n}\medskip\\
\frac{dR_{n}}{dt}=\gamma(A_{n}+I_{n})-\left( \varphi+\mu_{0}+r\right)
R_{n}\medskip\\
\frac{dQ_{n}}{dt}=r(N_{n}+I_{n})+\left( \Lambda-\mu_{0}\right) Q_{n}%
\medskip\\
N_{n}=S_{n}+A_{n}+R_{n}%
\end{array}
\right\}
\end{array}
\]%
\[%
\begin{array}
[c]{c}%
\text{Infectious population.}\medskip\\
I\!\left( t\right) =I_{c}\!\left( t\right) +I_{n}\!\left( t\right)
+\xi\left( A_{c}\!\left( t\right) +A_{n}\!\left( t\right) \right)
\medskip\\
J\!\left( t\right) =\frac{I\!\left( t-\tau_{p}-\theta_{p}\right) }%
{K_{I}+I\!\left( t-\tau_{p}-\theta_{p}\right) }\\
\\
\text{Variables to be fitted.}\medskip\\
\left\{
\begin{array}
[c]{l}%
\text{new cases}=(1-\rho)\int_{t-\Delta t}^{t}\beta\!\left( \mathfrak{t}%
\right) S\!\left( \mathfrak{t}\right) J\!\left( \mathfrak{t}\right)
d\mathfrak{t}\medskip\\
\text{total cases}=(1-\rho)\int_{0}^{t}\beta\!\left( \mathfrak{t}\right)
S\!\left( \mathfrak{t}\right) J\!\left( \mathfrak{t}\right) d\mathfrak{t}%
\end{array}
\right\}
\end{array}
\]
$\medskip$

\subsection{Environmental Effects}

Significant rain events will cause overflowing of river and stream banks,
which we assume will increase direct contact with the bacteria via contaminated
water, or indirectly via soils, vegetation and pathogen-carrying insects,
\emph{etc.}, that have been contaminated or have consumed cholera bacteria. We also
assume that temperature plays a role by increasing the infection rate. Warmer
temperatures result in more rapid growth, better survival and more active
bacteria and transmission agents. People also more frequently contact sources
of contamination due to the variety of increased activities that warmer
weather engenders.

Additionally, in the Ouest region, there are commercial areas, tent-cities and
slums that have raw waste directly discharging to Port-au-Prince Bay or to
the bay via rivers (e.g., Froide, Momance and Grise), as well as numerous open sewage
canals. \ We hypothesize that there may be additional disease generated when
large tidal ranges stir up contaminated sediments or cause blooms of plankton
in these coastal or estuarine waters. (We tested tidal height and found no association.)\ Again, the chain of infection may be complex and include contact with water, insects, plankton, benthos or
consumption of contaminated seafood, \emph{ etc}. \cite{deMagny, Huq, Fotedar} \ Why
tidal range would have a significant effect and not tidal height is curious.
There are a number of possible explanations. It may be that bottom sediments
possibly contain cholera biofilms \cite{Charles} or copepods hosts. A greater
tidal range would scour a larger bottom area by the breaking surface waters
along the beaches, or more fresh and brackish water from river outflows and
estuaries can contact the bottom sediments as the water falls and rises again,
or a combination of actions. In any event, these are just speculations, and the
association noted here may warrant closer examination in the field.

To incorporate these effects, we modify the transmission rate, $\beta$, as
follows:
\begin{equation}
\beta\!\left( t\right) =\alpha\left[ H\!\left( t\right) P\!\left(
t-\tau_{p},\theta_{p}\right) +cM\!\left( t-\tau_{m},\theta_{m}\right)
\right] \label{betaequ}%
\end{equation}
where $P$ is a moving average of the amount of daily rainfall, $M$ is a
moving average of the maximum semi-diurnal tidal range, $H$ is a heat index
based on mean air temperature, $\tau_{p}$ is the lag time for precipitation,
$\tau_{m}$ is the lag time for tides, $\theta_{p},\text{and }\theta_{m}$ are
the respective averaging periods, $\alpha$ is a proportionality constants and
$\text{and }c$ is the strength of the tidal effect relative to precipitation.

The population is split into two sub-populations. The first (core) group includes the worst slums and permanently displaced people living in tent camps or other privations. The other (non-core) group includes all others, from those that may have potentially been at risk only immediately after the earthquake to those that are poor, but have been fortunate enough to, through their own efforts or aid from outside agencies, see continual, albeit incremental, improvements in conditions over time. We include an expression for the improvement of conditions over time by having the number of susceptible or at-risk people reduced directly by public
health improvements. \ This improvement could be due to decreasing the number
of people in temporary housing, increased access to clean water, increased
personal hygiene, decreased contamination of the environment and rapid
treatment of new cases; or any other means of removing risk. \ We model this
by assuming that there is a constant rate of removal of people from the
exposed non-core population. We denote $r$ as the rate of improvement in
conditions; $1-u_{0}$ is the initial fraction of the population whose risk is
eradicable, and $u_{0}$ is the initial fraction of the population that is
chronically indigent. The at-risk population starts out equal to the entire
susceptible population and declines asymptotically to the remaining number of
indigent susceptibles as \linebreak conditions improve.

\subsection{Data}
\vspace{-12pt}

\subsubsection{Cholera Cases}

The source of the epidemic data was the Haitian Ministry of Public
Health and Population \cite{MSPP} and compiled by the Pan American
Health Organization \cite{PAHO}. The available data sets from the above
source, used for this study, consist of cumulative cholera cases and
new cholera cases. New cholera cases are calculated based on the
difference between the latest report and the previous one. The cholera case
definition also includes suspected cholera cases and deaths in addition to
confirmed cases and deaths. As such, data posted on the website are
periodically updated with minor corrections. Cases are reported on a weekly
basis, with the reporting week beginning on Sunday.

Reported hospitalized cases and hospitalized deaths are probably more
accurate, but for the purposes of this study, less useful, since we want to
track the progress of the epidemic, and increases in access to
treatment biases the data. \ No data is available for new or total cases
$\left( TC\right) $ during the first four weeks; however, hospitalized case
data is available. In order to estimate the number of new and total cases
during the first four weeks of the epidemic, we calculated the rate of increase
of the reported hospitalized cases over the first four weeks and then used
these rates to back-calculate the number of cases during the the first four
weeks of the epidemic necessary to match the reported numbers for all cases in
the fifth week.

\subsubsection{Environmental Data}

We compared rainfall, temperature and tide data to the pattern of new cholera
cases reported each week. The rainfall data comes from NASA \cite{NASA}, using
centrally-located points in each region as our datum point, given by
(latitude, longitude). For Artibonite, our datum point is
($19.125$,$~-72.625$) and for Ouest, it is ($18.625$,$~-72.375$) \cite{NASA}.
Precipitation estimates provided in the TRMM\_3B42\_daily.007 data product are
a combination of remote sensing and ground verified information reported with
a spatial resolution of $0.25\times0.25$ degrees and a temporal resolution of
1 day; further details are available on the website \cite{NASA}.

Temperature data is mean daily air temperature at Port-au-Prince and is
reported by the Weather Underground \cite{wunder}. We used a sine function fit
to the annual cycle in simulations. This has the advantage of allowing us to
extrapolate temperature patterns for model projections, while at the same time not severely impacting model performance by including more raw data in the simulations.\linebreak Unfortunately, only temperatures from Port-au-Prince were available. The temperature index derived from Port-au-Prince data were used for both Ouest and Artibonite.

Tide data is for Port-au-Prince Bay (StationId: TEC4709) \cite{NOAA}. These
numbers are predictions from NOAA's tide model for this location and
are not direct measurements. NOAA's web site allows one to download
tide numbers for any date starting from 2010 and extending through 2014.
Again, only data from Port-au-Prince is available. The use of model
predictions rather than actual measurements may result in underestimation of
the effects during extreme events, such as tropical storms.

\subsubsection{Data Analysis}

The initial modeling was done by comparing data sets in the frequency
(Fourier) domain for new cases and rainfall in order to find any suggestions
of matching periodicity and/or time-lags. After that code was written in
Berkeley Madonna to simulate the dynamical system and study the environmental
data in order to match predictions to data (when we discuss the output
of our model, we will use the term ``prediction'' to indicate the numbers
generated during the calibration phase of the model and ``projection'' to
indicate the extrapolation of the model's past cases used for the calibration). Rainfall data was available only to the week of 27 January 2013 (Week 119), because
of the lag between the timing of rainfall events and what had been processed
at the time we accessed it \cite{NASA}.

\subsection{Model Calibration}

All modeling and calibration was done in Berkeley Madonna. We used values (or averages of ranges) for all demographic variables, initial values and the three model independent parameters for cholera (fraction asymptomatic, recovery rate and loss of immunity rate) as established by previously published research ($\rho$, $\gamma$, $\varphi$, $\Lambda$, $\mu_{0}$, $N_{0}$, $S_{0}$, $I_{0}$,
$A_{0}$, $R_{0}$, $D_{0}$, and $t_{0}$; see Table \ref{Table 1}). Since our
model contains several fitted parameters ($\theta_{p}$, $\tau_{p}$, $k$,
$\alpha$, $\theta_{m}$, $\tau_{m}$, $c$, $M_{0}$, $K_{I}$, $r$ and $u_{0}$;
see Table \ref{Table 2}), we needed to supply a plausible range and initial
value for each parameter in order to efficiently search the parameter space
for a best fit. These ranges and initial values were chosen by visually
fitting the new case output of the model to reported new cases. \ The
Berkeley Madonna curve fitting algorithm was then used to minimize the root
mean square difference between model predictions of the number of new cases
for the time delays and averaging intervals ($\tau_{p}$, $\tau_{m}$,
$\theta_{p}$, $\theta_{m}$); these were then fixed, and the number of
cumulative cases was fit to reported cumulative cases by adjusting the
other parameters ($k$, $\alpha$, $c$, $M_{0}$, $K_{I}$, $r$ and $u_{0}$).
Parameter sensitivities were output from Berkeley Madonna at one-week intervals.
\ Confidence and prediction intervals were then calculated using the delta
method. \cite{Banks, Ramsay}\ For example, confidence intervals for the parameter estimates were obtained by inverting the self product of the sensitivity matrix and using the result to estimate the covariance matrix of the model parameters \cite{Banks}. These errors are only errors from the fitting procedure and do not include the propagation of error from the data themselves (which \linebreak are unknown).


\subsection{Parameters from Literature}

Table \ref{Table 1} displays parameter values for each region obtained from
published or online sources. The initial population distribution among core or
non-core sub-populations was obtained by multiplying each compartment by
$u_{0}$ or $1-u_{0}$, respectively.


\subsection{Environmental Components of the Force of Infection}
\vspace{-12pt}

\subsubsection{Artibonite}

Infections follow a key set of weather and climatic variables. Meteorological
influences on the infection rate are a product of the precipitation rate and a heat
index. Daily precipitation $p\left( t\right) $ \cite{NASA} is averaged
over an interval $\theta_{p}$, so that the running average precipitation rate is:
\[
P\!\left( t-\tau_{p},\theta_{p}\right) =\frac{1}{\theta_{p}}\sum
_{j=0}^{\theta_{p}}p\left( t-\tau_{p}-j\right)
\]
where $\tau_{p}$ is the delay in the precipitation's affect on the infection rate
(see below). For the delay, we used a simple sine function to model a mean
temperature index throughout the year (temperature data \linebreak from \cite{wunder}).
\ The mean air temperature index is given by:
\[
T_{air}\left( t\right) =\sin\left( \frac{2\pi\left( 7t+187\right)
}{365.25}\right) ,\text{ \ \ \ \ where }t\text{ is in weeks from
17 October 2010}
\]

In addition, there is a direct influence of temperature on the infection rate.
This would be expected, since warmer temperatures mean faster growth rates for
bacteria and some of their invertebrate hosts (bacterial dormancy is probably
not an issue, since the climate is tropical) \cite{deMagny, Huq}.\ We use a
heat index, $H\left( t\right) $, rather than temperature itself. This heat
index is a linear function of the normalized temperature pattern with a mean
(intercept) of $1$, and the slope, $k$, is a parameter to be fit. This index is
used as a multiplicative factor modifying the infection rate: the mean
temperature has no effect on the infection rate; low temperatures decrease the
infection rate, and high temperatures increase it. Thus, we have:
\[
H\left( t\right) =1+kT_{air}\left( t\right)
\]
Therefore, the infection rate is given by:
\begin{equation}
\beta_{A}\left( t\right) =\alpha H\!\left( t\right) P\!\left( t-\tau
_{p},\theta_{p}\right)
\end{equation}

The tide term, $\alpha cM\!\left( t-\tau_{m},\theta_{m}\right) $, is not
included, since we had no tidal range data for the Artibonite coast, and the
tidal range data we had (Port-au-Prince) was not found to explain a
significant amount of variance in the number of cases in Artibonite.

\subsubsection{Ouest}

For the Ouest region, we use the same formulation as in Artibonite; however, we
found that tidal range appeared to significantly affect infection
rates, as well. \ The maximum tidal range each day (there are two) $m\left(
t\right) $ \cite{NOAA} is averaged over an interval $\theta_{m}$, so the tidal
range formula is:
\[
M\!\left( t-\tau_{m},\theta_{m}\right) =\max\left\{ \left( \frac{1}%
{\theta_{m}}\sum_{j=0}^{\theta_{m}}m\left( t-\tau_{m}-j\right) \right)
-M_{0},0\right\}
\]
where $\tau_{m}$ is the delay in the tide's affect on the infection rate and
$M_{0}$ is the threshold for tidal range influence.\ \ Here, the effect of
water temperature on the response in infection rate was not found to be
sufficient to warrant adding another function and additional parameters.
\ Thus, the overall infection rate for Ouest is:%
\begin{equation}
\beta_{O}\left( t\right) =\alpha\left[ H\!\left( t\right) P\!\left(
t-\tau_{p},\theta_{p}\right) +cM\!\left( t-\tau_{m},\theta_{m}\right)
\right]
\end{equation}


\section{ Results}
\vspace{-12pt}

\subsection{Parameter Fitting and Model Selection}

Table \ref{Table 2} displays parameter values for each region obtained through
curve-fitting to cumulative reported cases.


Plausible ranges for time lags were initially obtained from the Fourier
analysis; then, parameter ranges and initial values were further refined by
visually fitting the new cases predicted by the model to the new case data.
We then used the Berkeley Madonna curve-fitting routine to find the remaining parameter set that minimized the sum of the square differences (SSD) between model output for cumulative cases and cumulative case data.

The half saturation constant $(K_{I})$ for the saturating infected function was very small compared to the numbers of infected people in Ouest and essentially zero in Artibonite (Table \ref{Table 2}). This indicates a very weak link between the numbers of infected and transmission rates. It may be that it requires only a handful of new cases to refresh the bacteria in the environment and/or there is a reservoir of viable bacteria in the environment itself (e.g., in plankton). In any event, we can force the dynamics of cholera in these two departments almost entirely by environmental conditions.

The Artibonite model with tide was not included in Table
\ref{Table 2}, since inclusion of tide only slightly improved the model fit
with cumulative cases and did not significantly improve the model fit for new
cases (see Tables \ref{Table 3} and \ref{Table 4}). The statistics for the
model fit are given in the following two tables. Table \ref{Table 3} is for
cumulative cases predicted by the model compared to cumulative case data.

\begin{table}[H]%
\centering
\small
\begin{tabular}
[c]{ccccc}\toprule

\multicolumn{5}{c}{\textbf{All Cases}}\\\midrule

\textbf{Statistic} & \textbf{Artibonite (Tide)} & \textbf{Artibonite (No Tide)} & \textbf{Ouest (Tide)} &\textbf{Ouest}
\textbf{(No Tide)}\\\midrule
Data points & $153$ & $153$ & $153$ & $153$\\
Parameters & $11$ & $7$ & $11$ & $7$\\
Adj RMSD & $1603.78$ & $1686.48$ & $2825.08$ & $3773.74$\\
Adj R$^{2}$ & $0.996$ & $0.995$ & $0.998$ & $0.997$\\\bottomrule

\end{tabular}

\caption{{ {Model predictions \emph{versus} data statistics for
the cumulative number of cases: root mean squared deviations (RMSD),
coefficient of determination $(R^{2})$. Degrees of freedom for the statistics
are adjusted by the number of parameters fit in the calibration process.}}}%
\label{Table 3}%
\end{table}

 \begin{table}[H]%
\centering
\small
\begin{tabular}
[c]{ccccc}\toprule

\multicolumn{5}{c}{\textbf{New Cases}}\\\midrule

\textbf{Statistic} & \textbf{Artibonite (Tide) }&\textbf{Artibonite (No Tide)} & \textbf{Ouest (Tide)} & \textbf{Ouest
(No Tide)}\\\midrule
Data points & $153$ & $153$ & $153$ & $153$\\
Parameters & $11$ & $7$ & $11$ & $7$\\
Adj RMSD & $579.79$ & $587.09$ & $1702.26$ & $1868.38$\\
Adj R$^{2}$ & $0.785$ & $0.789$ & $0.476$ & $0.488$\\\bottomrule

\end{tabular}

\caption{{ {Model predictions \emph{versus} data statistics for
new cases: root mean squared deviations (RMSD), coefficient of
determination $(R^{2})$. Degrees of freedom for the statistics are adjusted by
the number of parameters fit in the calibration process.}}}%
\label{Table 4}%
\end{table}


For the full model in either region (model including tides), the parameters
$\theta_{m},\tau_{m},M_{0}$ and $c$ are added and the\ model is re-optimized.
\ \ For Artibonite, an $F$-test for the nested models with cumulative cases
gives the following results $F=$\ $4.86$ \ $d.f.=\left( 4,\;142\right) $, \ and
the $p$-value is $0.001$. \ For the Ouest region, an $F$-test for the nested
models gives the following results $F=$\ $29.63$ \ $d.f.=\left( 4,\;142\right)
$, \ and the $p$-value is $6.46\times10^{-18},$ indicating that the tidal data
significantly improved the model fit.

Table \ref{Table 4} is for new cases predicted by the model compared to new
case data. The $F$-test for the Artibonite nested models
using new cases gives the following results $F=$\ $1.93$ \ $d.f.=\left(
4,\;142\right) $, \ and the $p$-value is $0.109.$ For Ouest, the difference is
again significant with $F=$\ $8.47$ \ $d.f.=\left( 4,\;142\right) $, \ and the
\linebreak$p$-value is $3.698\times10^{-6}.$


\subsection{Lag Times}

The total delays in response to precipitation and tides are the sum of the
averaging window and the delay function. For precipitation in Artibonite, the
averaging window is five weeks plus a $3.4$-week delay for a total delay range
of $3.4$ to $8.4$, and in Ouest, they are $2.3$ and $5.1$ weeks, respectively,
for a total delay range of $5.1$ to $7.4$ weeks. Thus, the delays are similar
in the two regions. These long delays are similar in magnitude to delays
reported from a study of cholera in Zanzibar, East Africa (eight-week delay)
\cite{Reyburn}, but slightly longer than those reported in Bangladesh (four
weeks) \cite{deMagny} and much longer than reported in another study of the
Haiti cholera epidemic (four to seven days) \cite{Eisenberg}. \ For Ouest, the estimated
averaging window and delay from response to changes in tidal range were about
one week and two weeks, respectively, or a delay range of one to three weeks total.
However, since the influence of tidal range has not been quantitatively reported
elsewhere in the literature, we have nothing with which to compare this number.

At the time of this analysis, rainfall data were available only to the week of
29 September 2013 \linebreak(Week 154, with the lag periods reported above,\ this brings the
simulation out to 27 November for Artibonite and 20 November for Ouest. This is just
before the data for new cases ends on 8 \linebreak December 2013.


\subsection{Vaccination}

A program to vaccinate the most at risk populations began in the second
week of April and ended in mid-June, 2012. Each site (Ouest and Artibonite
Departments) vaccinated about 50,000 persons, and each site had about a $91\%$ second dose
coverage. The administration of the first dose was staggered by age group
(beginning first with 10-year-olds and up), because the Ministry of Health had
a measles, rubella and polio vaccine catch-up campaign for children under 10
years of age that was taking place at the same time last April
2012 (communicated by Jordan Tappero, MD, MPH (CDC/CGH/DGDDER,
Atlanta, GA, USA), 29 November 2012).

In the Ouest Department, GHESKIO (Groupe Ha\"{\i}tien d'\'{E}tude du
Sarcome de Kaposi et des Infectieuses Opportunistes) vaccinated adults,
adolescents and children over 10 years of age from 12--23 April 2012, and
children under 10 from 26 May--3 June. The first dose of vaccine was given to
52,357 persons (of which, 47,520 received the second dose), living in the slums
of Port-au-Prince and surrounding villages (communicated by Jean W.
Pape, MD (GHESKIO, Weill Cornell Medical College, Port-au-Prince, Haiti), 29 November 2012).

In the Artibonite Department, PIH (Partners in Health) vaccinated
32,183 people in rural Bocozel and 13,185 people in Grand Saline, with $90.8\% $
of those people confirmed to get the second dose (or 41,194 for both locations).
The campaign started 15 April 2012, and ran until 10 June 2012. Here, too,
children under nine years old were vaccinated in the second half of the time
period, because of the MMR and Polio vaccination
campaign (communicated by Louise Ivers, MD (Partners in Health/ZL,
Cange, Haiti), 29 November 2012).

With these basic facts, we constructed a crude vaccination schedule (Table
\ref{Table 5}) using the\linebreak following assumptions:

(1) approximately $25\%$ of the population is under 10 years old;

(2) the second dose was administered 14 days after the first dose was given
\cite{Date};

(3) the immune response took hold about 8.5 days after the second dose was
given \cite{Date};

(4) we used the average number of people vaccinated per day over a 12-day
period for adults and \linebreak 9 days for children.

 \begin{table}[H]%
\centering
\small
\begin{tabular}
[c]{lll}\toprule

\multicolumn{3}{c}{$\text{\textbf{Ouest}}$}\\\midrule

$\text{Adult }1{st}\text{ dose }$ & $39,268$ & $\text{10 April--23 April}%
$\\
$\text{Adult }2{st}\text{ dose}$ & $35,640$ & $\text{26 April--7 May}%
$\\
$\text{Adult immune response}$ & $65\%$--$85\%$ & $\text{4 May--15 May}%
$\\
$\text{Child }1{st}\text{ dose }$ & $13,089$ & $\text{26 May--3 June}%
$\\
$\text{Child }2{st}\text{ dose }$ & $11,880$ & $\text{9 June--17 June}%
$\\
$\text{Child immune response}$ & $65\%$--$85\%$ & $\text{17 June--25 June}%
$\\
\end{tabular}
\begin{tabular}
[c]{lll}\midrule

\multicolumn{3}{c}{$\text{Artibonite}$}\\\midrule

$\text{Adult }1{st}\text{ dose }$ & $34,026$ & $\text{15 April--26 April}%
$\\
$\text{Adult }2{st}\text{ dose}$ & $30,896$ & $\text{29 April--10 May }%
$\\
$\text{Adult immune response}$ & $65\%$--$85\%$ & $\text{7 May--18 May}%
$\\
$\text{Child }1{st}\text{ dose }$ & $11,342$ & $\text{19 May--27 May}%
$\\
$\text{Child }2{st}\text{ dose }$ & $10,298$ & $\text{2 June--10 June}%
$\\
$\text{Child immune response}$ & $65\%$--$85\%$ & $\text{10 June--18 June }%
$\\\bottomrule

\end{tabular}

\caption{{ {Simulated vaccination schedule for Artibonite and
Ouest.}}}%
\label{Table 5}%
\end{table}

\medskip

We ran simulations following the above schedule as closely as the simulation
would allow by subtracting the numbers of at-risk persons (given below) from
the susceptible compartment $(S)$.

In Ouest, the vaccination algorithm involved removing 2970 at-risk persons per
day starting\linebreak on 4 May 2012, and ending on 15 May 2012, or 35,640 total (these
correspond to the vaccination of persons over age 10). Then, the algorithm
removed 1320 at-risk persons per day starting on 17 June 2012, and ending on
25 June 2012, or 11,880 total (children).

In Artibonite, the vaccination algorithm removed 2574.67 at-risk persons per
day starting\linebreak on 7 May 2012, and ending on 18 May 2012, or 30,896 total (over age
10); then, the algorithm removed 1144.22 persons per day starting on
10 June 2012, and ending on 18 June 2012, or 10,298 total (these correspond to
the vaccination of the at-risk children). These simulations roughly follow the
actual vaccination schedules given in Table \ref{Table 5}.

The efficacy of the vaccine (oral Shanchol) is reported to be between $65\%$ and
$85\%$ \cite{Date}. We fit efficacy as a parameter and found for Artibonite an
efficacy of $67.5\%$ and for Ouest $70.3\%$.


\subsection{Simulations and Projections}

We list the time line for particular events in Table \ref{Table 6}. Curve
fitting (parameter estimation) was done between model output and data from
Week 3 to Week 155. We refer to model output during the calibration period as
\textquotedblleft predictions''; simulations from Week 156 through Week 216 are
referred to as \textquotedblleft projections''. We used rainfall data that
ended in Week 154 for the delay period. Data for new infections extended to
Week 164. By \textquotedblleft immune response for first...'' and
\textquotedblleft immune response for last...'', we mean that this is when we begin
and end removing susceptibles from the at-risk group, respectively.

\begin{table}[H]
\centering
\small
\begin{tabular}
[c]{ccc}\toprule

& \multicolumn{2}{c}{\textbf{Date (Week of Epidemic)}}\\\cline{2-3}%
\textbf{Event} & \textbf{Artibonite} & \textbf{Ouest}\\\midrule

{\small Begin epidemic} & \multicolumn{2}{c}{{\small 17 October 2010 \ (0)}%
}\\
{\small Begin case data} & \multicolumn{2}{c}{{\small 7 November 2010 (3)}}\\
{\small End model fitting} & \multicolumn{2}{c}{{\small 6 October 2012 (155)}%
}\\\midrule

{\small Immune response for first adult vaccinated} & {\small 7 May 2012 (81.1)}
& {\small 4 May 2012 (80.7)}\\
{\small Immune response for last adult vaccinated} & {\small 18 May 2012 (82.7)}
& {\small 15 May 2012 (82.3)}\\\midrule
{\small Immune response for first child vaccinated} & {\small 10 June 2012 (86)}
& {\small 17 June 2012 (87)}\\
{\small Immune response for last child vaccinated} & {\small 18 June 2012 (87.1)}
& {\small 25 June 2012 (88.1)}\\\midrule
{\small End precipitation data} & \multicolumn{2}{c}{{\small 29 September 2013
(154)}}\\\midrule
{\small End precipitation data with delay} & {\small 27 November 2013 (162.4)} &
{\small 20 November 2013 (161.4)}\\

{\small Begin random average rain fall data} & {\small 28 November 2013 (162.5)} &
{\small 21 November 2013 (161.5)}\\\midrule
{\small End case data} & \multicolumn{2}{c}{{\small 8 December 2013 (164)}}\\
{\small End simulation} & \multicolumn{2}{c}{{\small 7 December 2014 (216)}%
}\\\bottomrule

\end{tabular}

\caption{Time line for modeling events.}%
\label{Table 6}%
\end{table}

\subsubsection{Predictions Compared to Observations for Cumulative and New
Cases}

The following Figures \ref{Fig1} and \ref{Fig2} show the model predictions
compared to observations for the cumulative number of cases in the Artibonite and
Ouest regions, respectively. Similarly, Figures \ref{Fig3} and \ref{Fig4} show
the model predictions compared to observations for the new number of cases in
Artibonite and Ouest regions, respectively. Prediction intervals were
calculated only for the cumulative numbers, since the final model fitting was
done on these numbers. Confidence intervals are shown for incidence. The match
for the trends in new cases match fairly well: the slope of the expected
(model) regressed against observed (data) is nearly one in both departments
(see Figures \ref{Fig5} and \ref{Fig6}), even though there is a substantial
amount of unexplained variance. Whether this is due to the crude spatial
resolution or other factors remains to be seen.

Prediction intervals (PIs) on the cumulative number of cases were calculated
using the delta method adapted for differential equations (see, for example,
Ramsay \emph{et al}. \cite{Ramsay}). After 29 September 2013, rainfall data from NASA was
unavailable; for simulations after that date, we did 13 runs using rainfall
patterns from each of the 13 prior years. The mean of those simulations was
used, and the variance of the 13 runs, at each time step, was added to the
variance from the estimation procedure before computing the PIs. We also
included the upper and lower 95\% percentiles of the rainfall patterns on projected
\linebreak new cases.
\begin{figure}[H]
\centering\includegraphics[width=7in,height=6in]{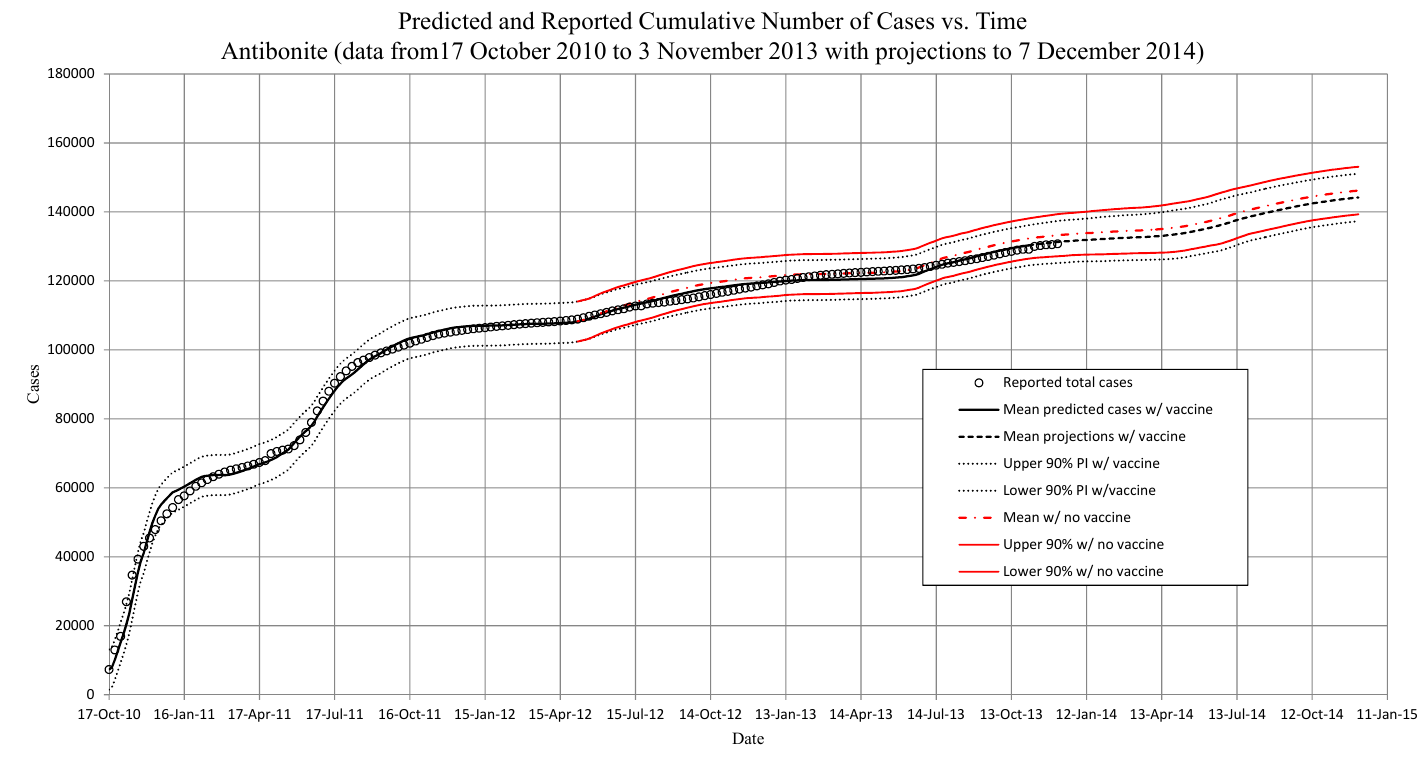}\caption{Artibonite.
The predicted cumulative number of symptomatic individuals, against
total reported cases to 1 April 2012. Projections are from then to the end of
February. Projections using the vaccination schedule (red) begin on
7 May 2012, approximately three weeks after beginning the vaccination program
in Artibonite. All projections after 11 November 2012, are based on runs using the prior
13 years of precipitation, and PI's include the variance of those data (see
the text).}%
\label{Fig1}%
\end{figure}
\begin{figure}[H]
\centering\includegraphics[width=7in,height=6in]{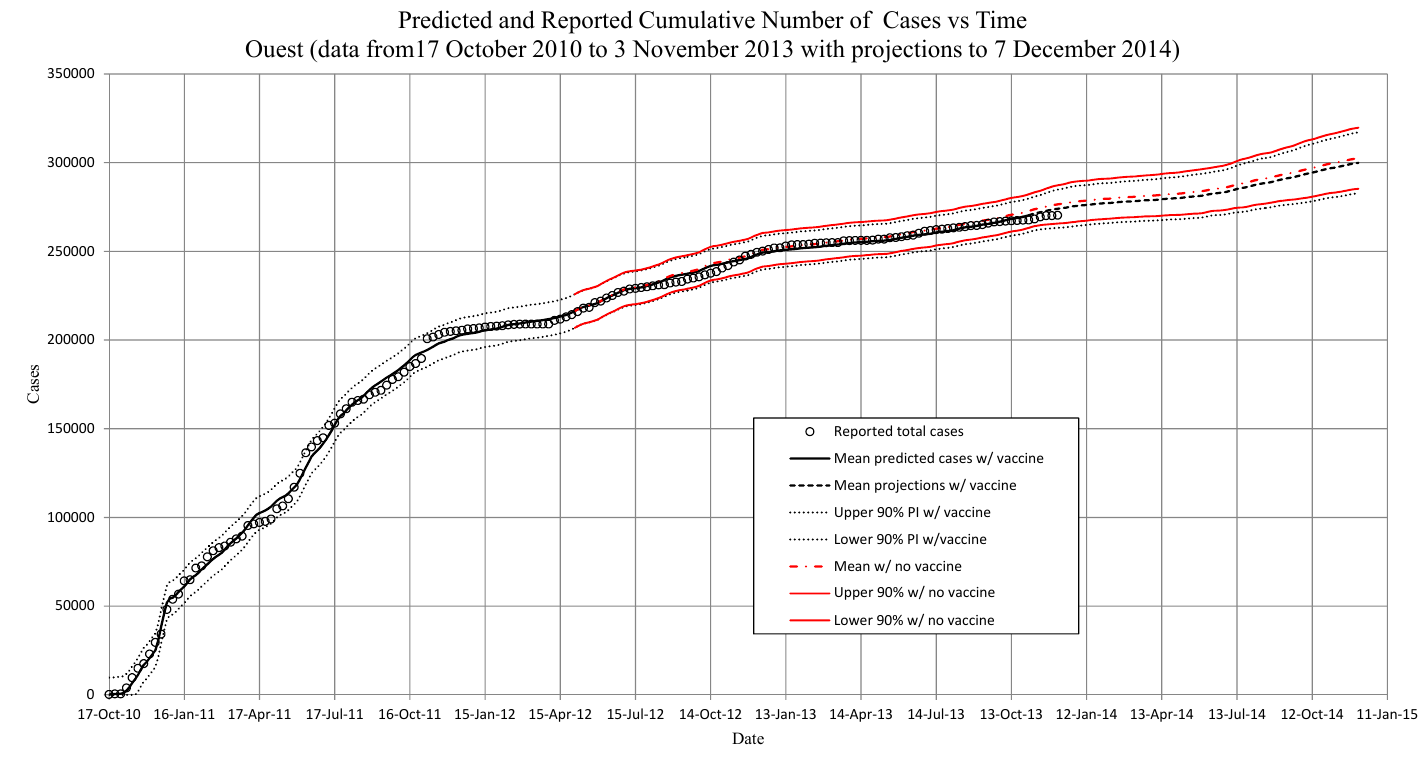}\caption{Ouest.
The predicted cumulative number of symptomatic individuals, against
total reported cases to 1 April 2012. Projections are from then to the end of
February. Projections using the vaccination schedule (red) begin on
4 May 2012, approximately three weeks after beginning the vaccination program
in Ouest. All projections after 11 November 2012, are based on runs using the prior 13
years of precipitation, and PI's include the variance of those data (see the
text). Note that for the Ouest region, the model begins at the fourth
week. We assume that the low initial numbers in the first three weeks are a
result of the immigration of cases from the Artibonite region. The model therefore
uses data for the first four weeks (assumed immigration numbers for the
first three weeks and the initialization of the model from data for the fourth
week).}%
\label{Fig2}%
\end{figure}
\begin{figure}[H]
\centering\includegraphics[width=7in,height=6in]{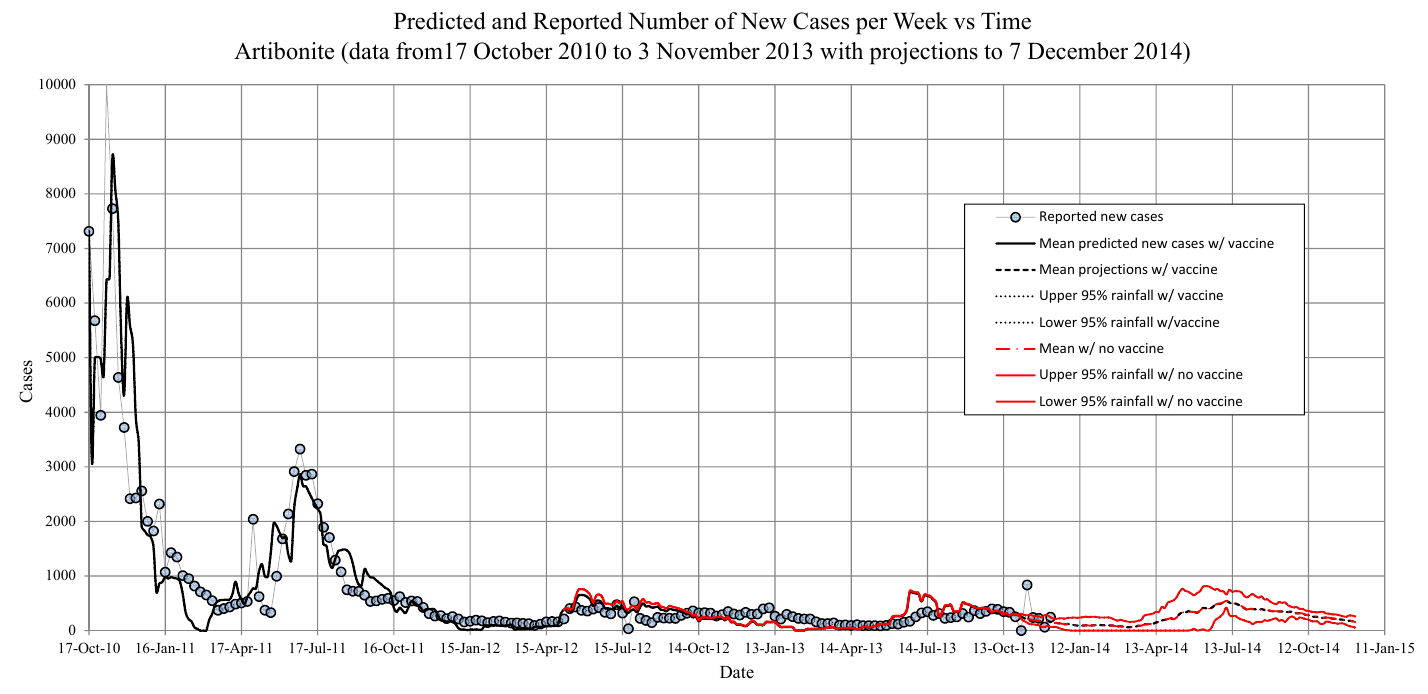}\caption{Artibonite.
The new symptomatic individuals \emph{vs}. time. Circles, observed; solid
line, model prediction; dashed lines, 95th percentile confidence intervals for
model projections. The red line is projections with 75 percent vaccine efficacy.}%
\label{Fig3}%
\end{figure}
\begin{figure}[H]
\centering\includegraphics[width=7in,height=6.5in]{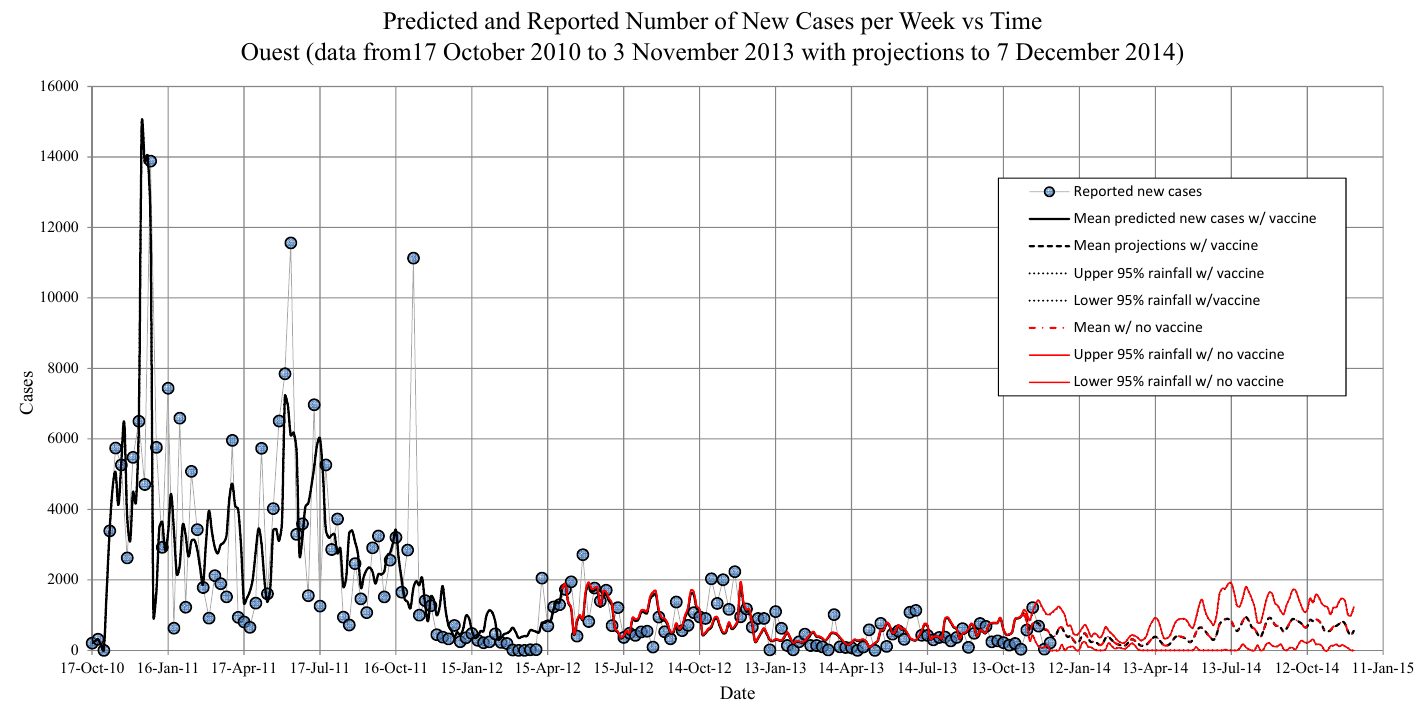}\caption{{ {Ouest.
The new symptomatic individuals \emph{vs}. time. Circles, observed; solid
line, model prediction; dashed lines, fifth, 50the and 95th percentiles for model
projections based on the past 13 years of precipitation records.}}}%
\label{Fig4}%
\end{figure}
\begin{figure}[H]
\centering\includegraphics
[width=7in,height=5in]{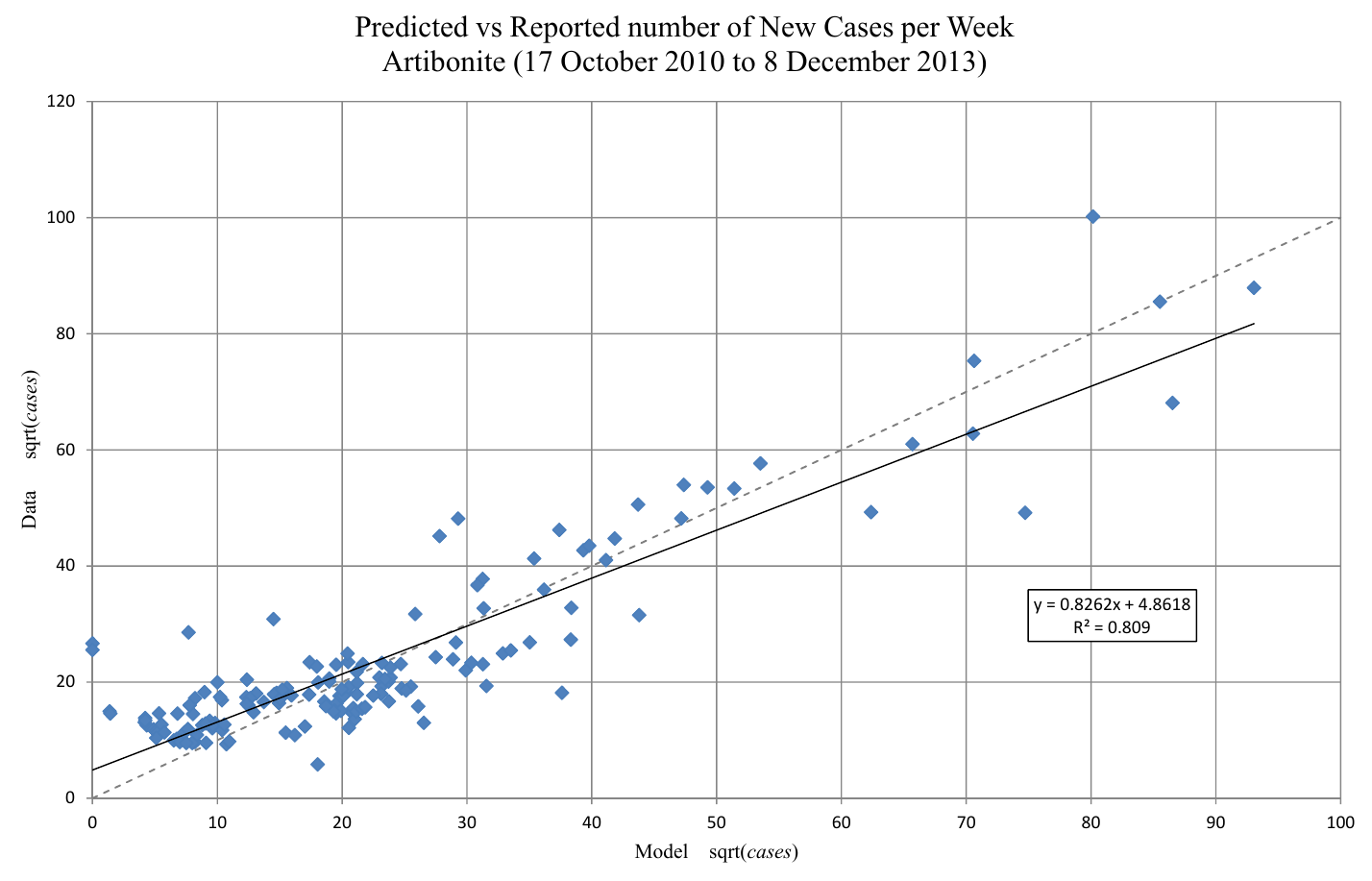}\caption{Artibonite.
The predicted new symptomatic individuals, against weekly reported
cases to 1 April 2012 (square root transformed). A regression line matching the
main diagonal $\left( 45^{\circ}\right) $ dashed line would show an optimal
fit; the discrepancy is due in part to fitting on the cumulative numbers.}%
\label{Fig5}%
\end{figure}

\subsubsection{Epidemic Predictions and Projections for Artibonite}

The model predicted that by 8 May 2013, one year after the vaccination
program, Artibonite would have seen between 117,000 and 128,000 cholera cases
without vaccine and between 115,000 and \linebreak 127,000 with the implemented
vaccination program (123,000 thousand actual), a decrease in about \linebreak 1700
cases (see Figure \ref{Fig7}).

According to the model without vaccine, an average of 11,989 people would have
gotten sick between 6 May 2012, and 7 November 2012 (six months), and over the same
interval with vaccine, an average of 10,375 people would have gotten sick.
This represents a $13.5\%$ reduction in the number of people that would have
gotten cholera. Between 6 May 2012, and 6 November 2013 (eighteen months), an
average of 24,188 people would have gotten sick without vaccine, and over the
same interval with vaccine, an average of 22,228 people would have gotten
sick. This represents an $8\%$ reduction in the number of people that would
have gotten cholera. The maximum percent reduction in the number of cases due to
vaccination occurs about 8 August 2012, $13.5$ weeks after the assumed beginning of
immunity with a $14.5\%$ reduction in cases. Percent reductions start to
taper off after this point, due to the loss of immunity, which is treated as an
exponential decay, and the continued occurrence of new cases, which eventually
dilutes the effect \linebreak(see Figure \ref{Fig8}).

\begin{figure}[H]
\centering\includegraphics[width=7in,height=6in]{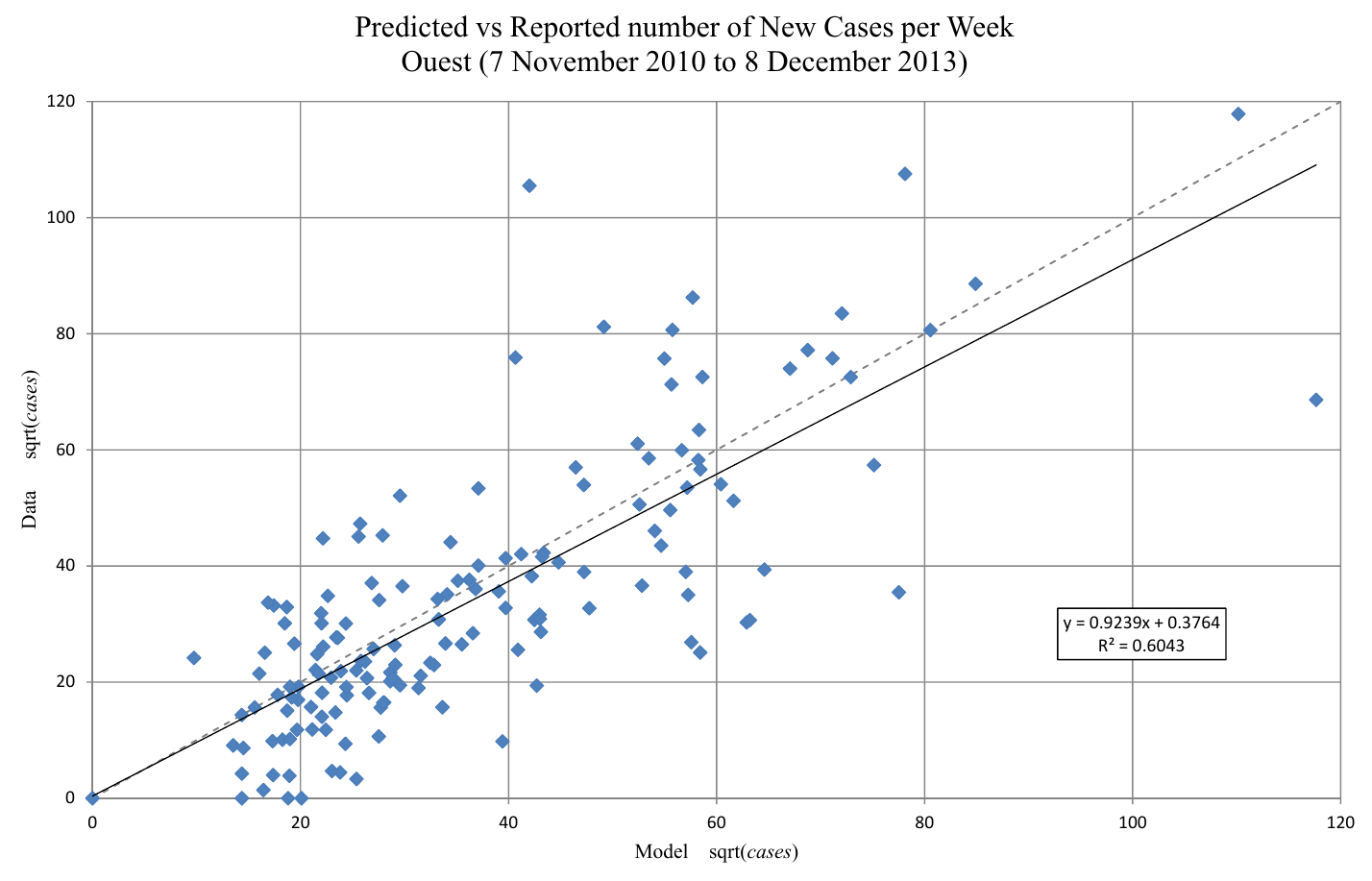}\caption{{ {Ouest.
The predicted new symptomatic individuals, against weekly reported
cases to 1 April 2012 (square root transformed). A regression line matching the
$45^{\circ}$ line would show an optimal fit, the discrepancy is due in part to
fitting on the cumulative numbers.}}}%
\label{Fig6}%
\end{figure}

\begin{figure}[H]
\centering\includegraphics[width=7in,height=6in]{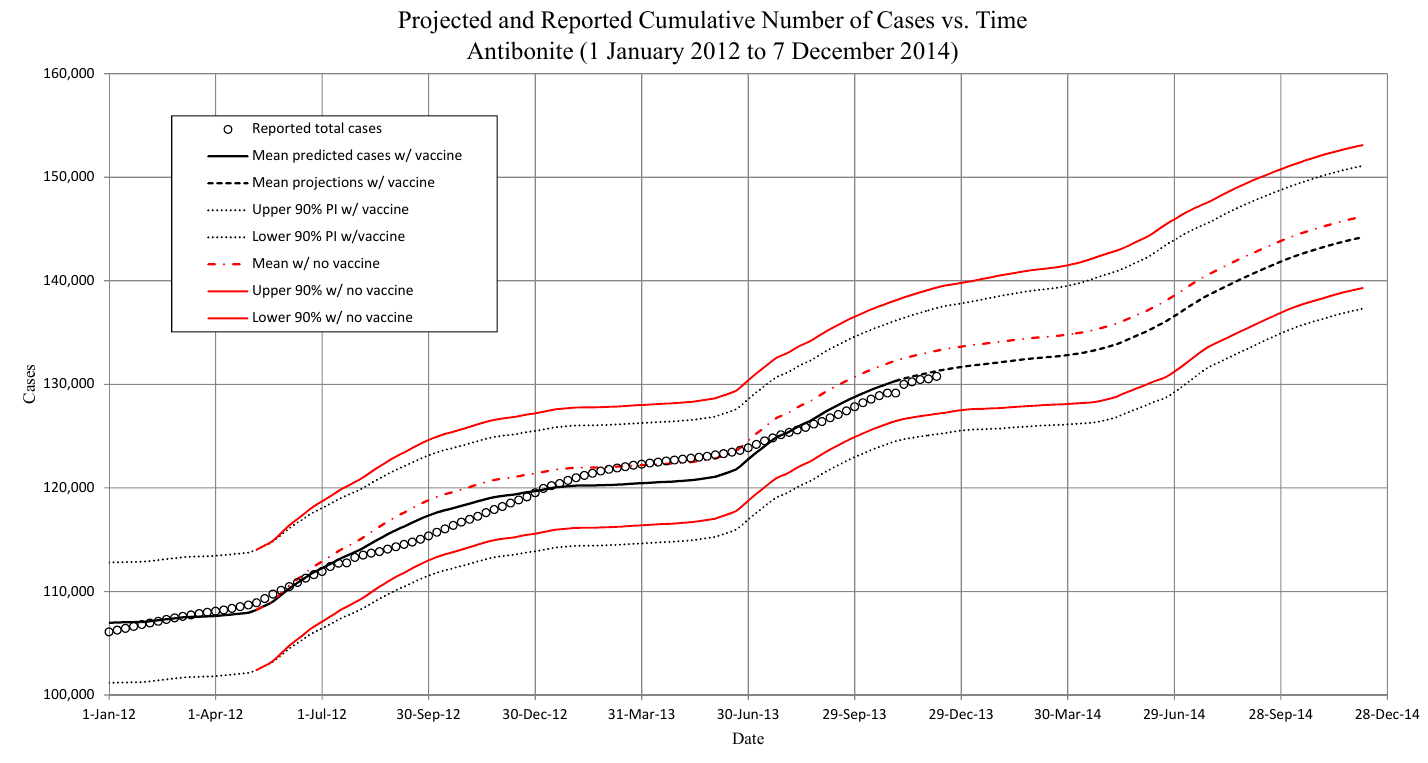}\caption{{ {\protect\scalebox{.95}[1.0]{Artibonite.
The projected total symptomatic individuals \emph{vs}. time. Circles, observed;} solid line, model prediction; dashed lines, fifth, 50th and 95th
percentiles for model projections based on the past 13 years of precipitation
records. The red lines are the model runs with the vaccination schedule that occurred in
spring, 2012 (see the text). }}}%
\label{Fig7}%
\end{figure}

\subsubsection{Epidemic Predictions and Projections for Ouest}

For Ouest, the model projected that by 5 May 2013, one year after the
vaccination program there, Ouest would have seen between 248,000 and 267,000
cholera cases without vaccine and between 246,000 and 265,000 with the
implemented vaccination program (257,000 actual), a decrease in about 1900 cases (see Figure \ref{Fig9}).

\begin{figure}[H]
\centering\includegraphics[width=7in,height=5in]{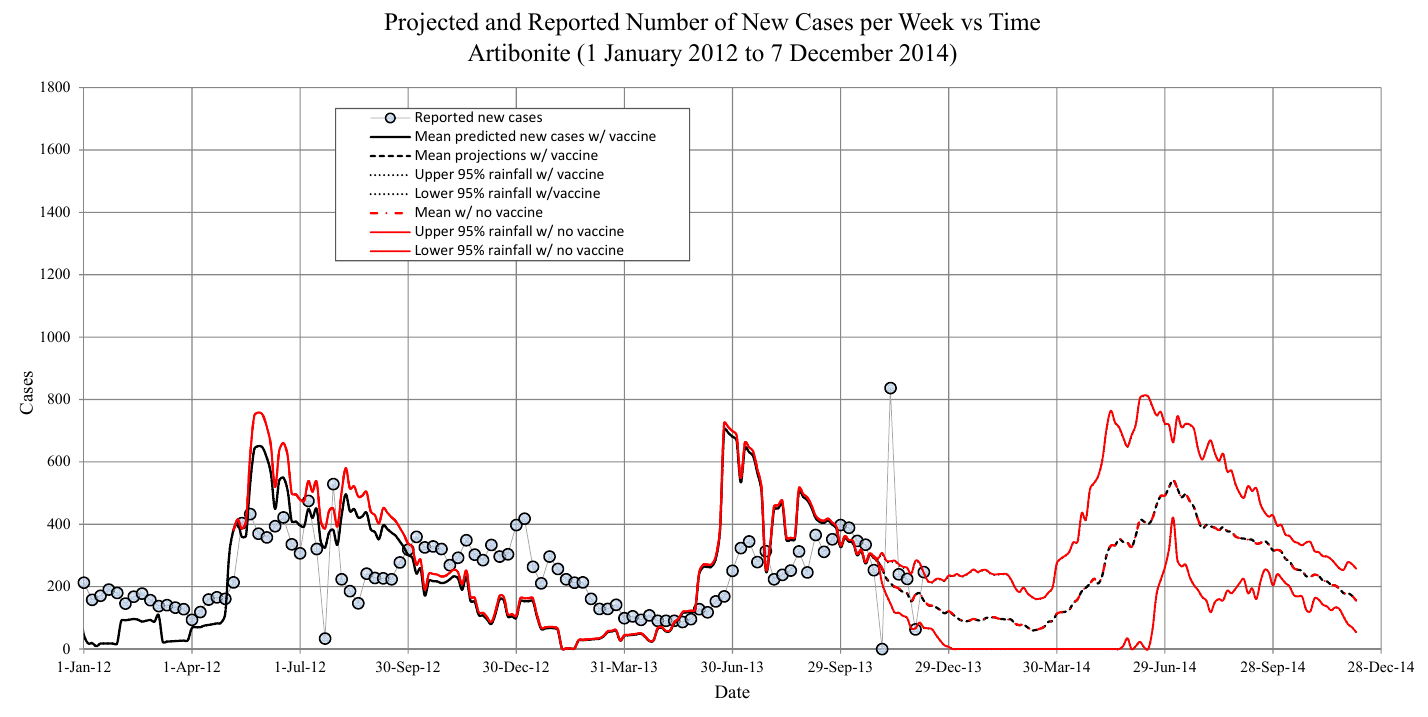}\caption{{ {\protect\scalebox{.95}[1.0]{Artibonite.
The projected new symptomatic individuals \emph{vs}. time. Circles, observed;} solid line, model prediction; dashed lines, 95the percentile
confidence intervals for model projections. The red line is projections with 75
percent vaccine efficacy.}}}%
\label{Fig8}%
\end{figure}

The Ouest model predicts, without vaccine, an average of 28,857 people would
have gotten sick between 2 May 2012, and 4 November 2012 (six months), and over the
same interval with vaccine, an average of 27,505 people would have gotten
sick. This represents a $4.7\%$ reduction in the number of people that would
have gotten cholera. \ Between 2 May 2012, and 3 November 2013 (eighteen months), an
average of 56,377 people would have gotten sick without vaccine, and over the
same interval with vaccine, an average of 54,003 people would have gotten
sick. This represents a $4.2\%$ reduction in the number of people that would
have gotten cholera. \ The maximum percent reduction in the number of cases due to
vaccination occurs about 21 November 2012, $29$ weeks after the assumed beginning of
immunity with a $4.74\%$ reduction in cases. \ The smaller fraction vaccinated
in Ouest also leads to much subtler differences in the incidence. \ Percent
reductions in Ouest start to taper off much later than in Artibonite, since the
fraction of the at-risk population protected starts out at a much lower level,
and the rate of decay of the effects are subsequently slower (see Figure
\ref{Fig10}).

\begin{figure}[H]
\centering\includegraphics[width=7in,height=7in]{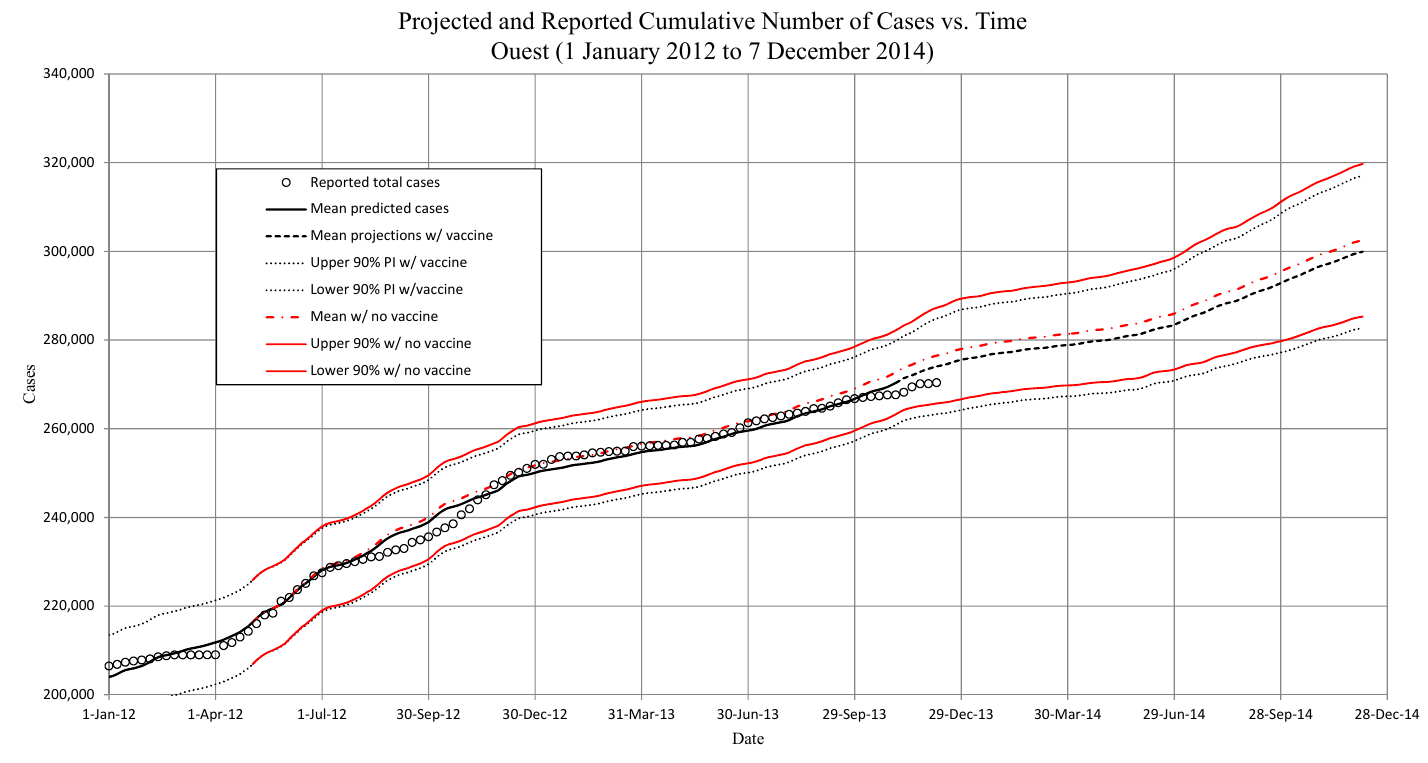}\caption{{ {Ouest.
The projected total symptomatic individuals \emph{vs}. time. Circles, observed; solid line, model prediction; dashed lines, fifth, 50th and 95th
percentiles for model projections based on the past 13 years of precipitation
records. The red lines are model runs with vaccination schedule that occurred in
spring, 2012 (see text). The slow down in growth in late February and March was
before the vaccination program began.}}}%
\label{Fig9}%
\end{figure}
\begin{figure}[H]
\centering\includegraphics[width=7in,height=6in]{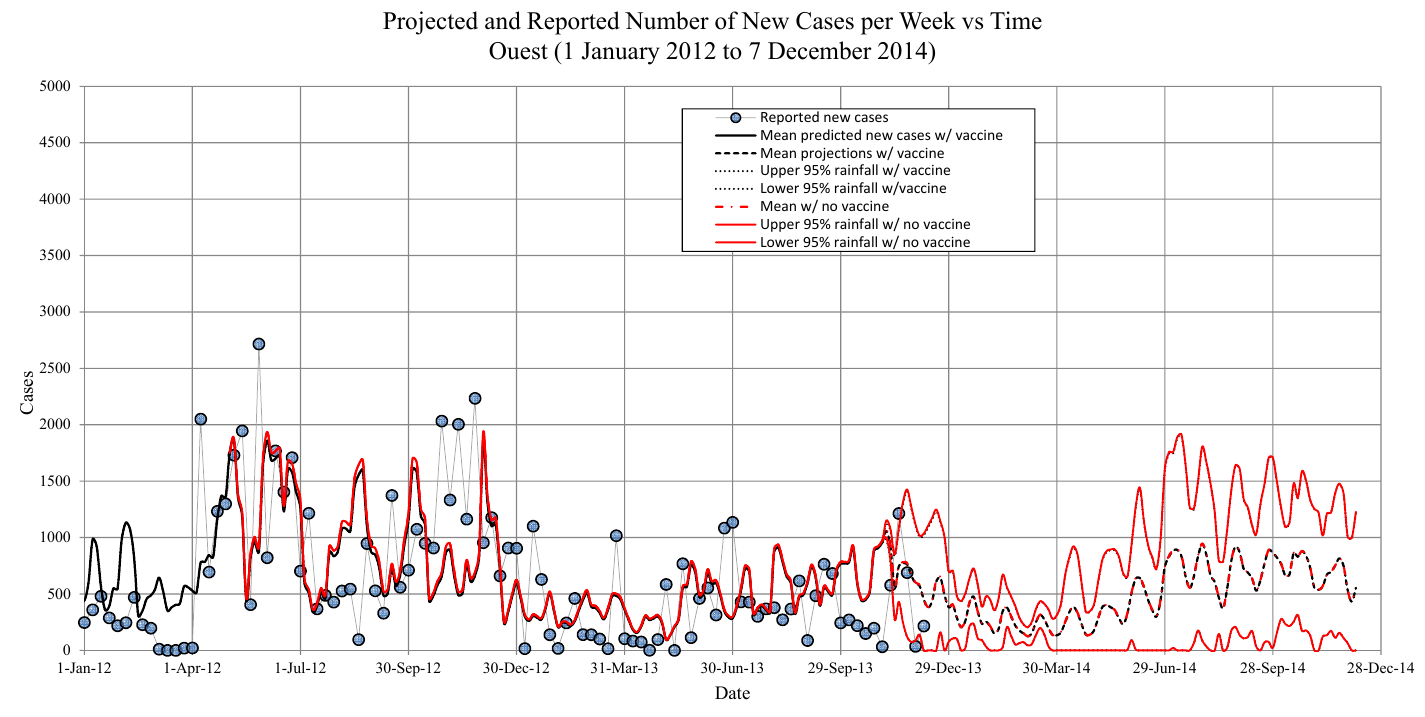}\caption{{ {Ouest.
The projected new symptomatic individuals \emph{vs}. time. Circles, observed; solid line, model prediction; dashed lines, fifth, 50th and 95th
percentiles for model projections based on the past 13 years of precipitation
records.}}}%
\label{Fig10}%
\end{figure}

\subsection{Vaccination Scenarios}

We looked at changing the number of people vaccinated and the timing of
vaccination to see if there is some optimal schedule that can be applied.

\subsubsection{Changing the Number of People Vaccinated}

\begin{figure}[H]
\centering\includegraphics[width=7in,height=6in]{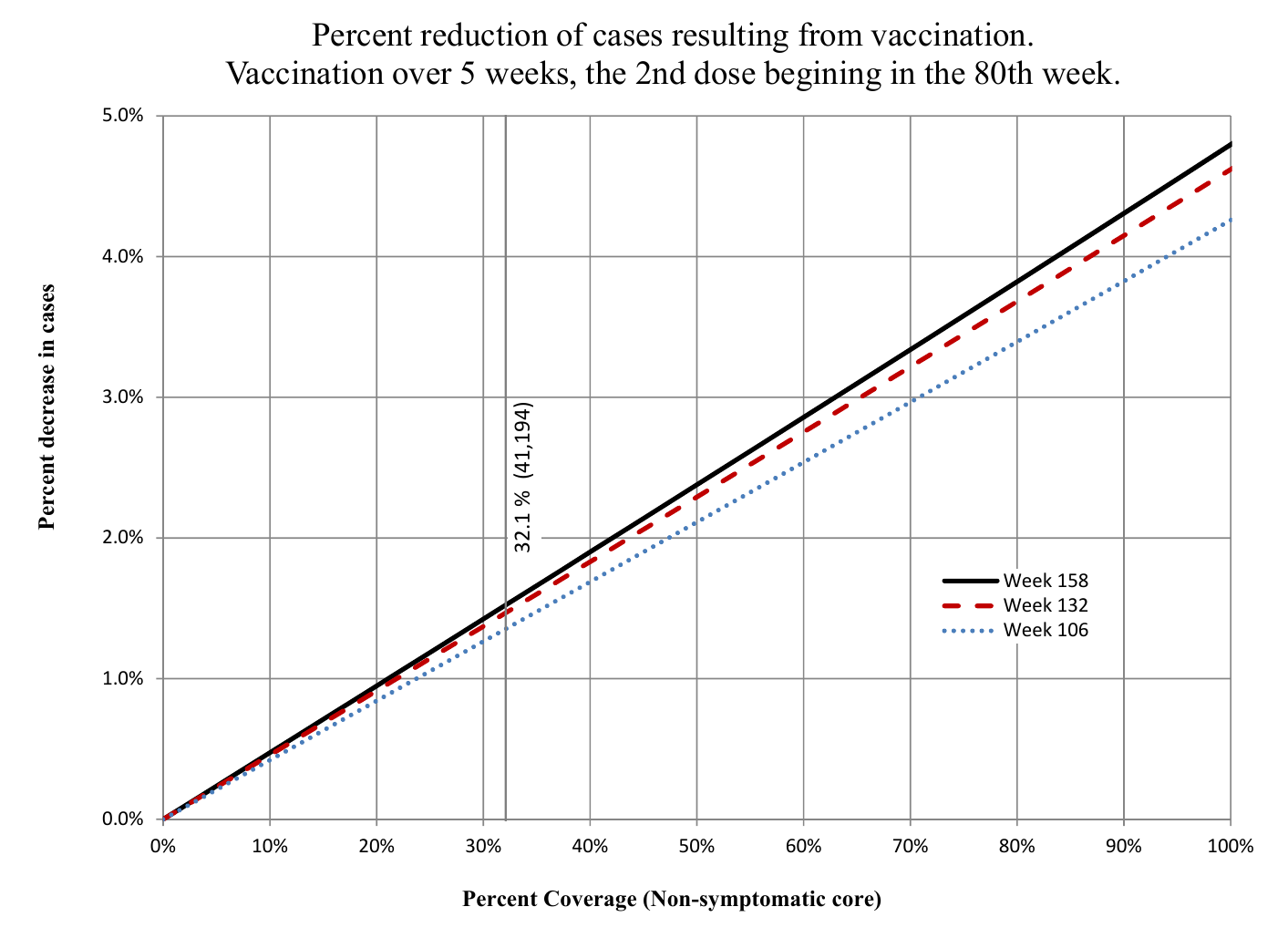}\caption{{ {Artibonite.
The percent decrease in the total number of cases at Weeks 100, 125 and 150
for the percent of the non-symptomatic core group vaccinated indicated on the
abscissa. The onset of the immune response was assumed to begin in Week 81 and
complete in Week 85. We assume $70\%$ efficacy of the vaccine. The actual
number completing vaccination (41,194) is indicated by the vertical line.}}}%
\label{Fig11}%
\end{figure}

\begin{figure}[H]
\centering\includegraphics[width=7in,height=6in]{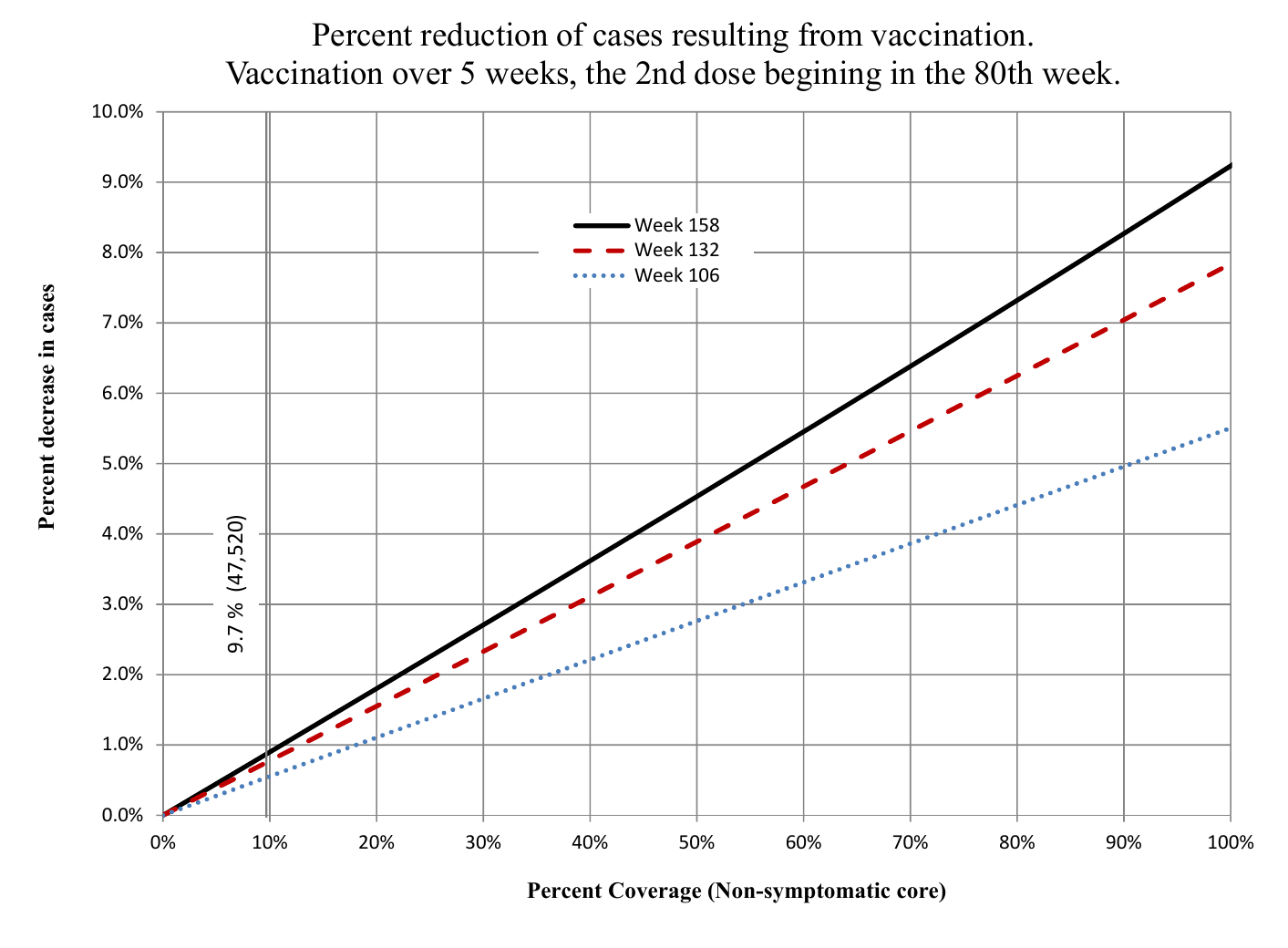}\caption{{ {Ouest.
The percent decrease in the total number of cases at Weeks 100, 125 and 150
for the percent of the non-symptomatic core group vaccinated indicated on the
abscissa. The onset of the immune response was assumed to begin in Week 81
and complete in Week 85. We assume $70\%$ efficacy of the vaccine. The actual
number completing vaccination (47,520) is indicated by the vertical line.}}}%
\label{Fig12}%
\end{figure}

The first experiment was to change the number of people vaccinated. We
completed, in the model, all vaccinations within a five-week period. The second
round of vaccination was assumed to begin in epidemiological Week 80, and the
immune response was assumed to begin a week later in Week 81, with a $70\%$
efficacy. This was to approximately match the timing of the initiation of the
actual vaccination second dose. We varied the numbers vaccinated from $0\%$ to
$100\%$ coverage of the non-symptomatic core group (susceptible, asymptomatic
and recovered asymptomatic), in Artibonite and Ouest. The results are
illustrated in Figures \ref{Fig11} and \ref{Fig12}. Percent coverage of the
vaccinated is shown on the x-axis and percent decrease in cases on the y-axis.
We show the percent decrease of cases at Weeks $100$, $125$ and $150$ of the
epidemic. These correspond to $20$, $45$ and $70$ weeks after beginning
administration of the second dose of the vaccine in the $80{th}$ week.


In both departments, the percent decrease in the number of cases increases steadily
until the number vaccinated reaches the number of people remaining in the at-risk group. At this point, there are no more people to be vaccinated, but $30\%$
of those people that were susceptible and vaccinated are still susceptible.

The curve increases almost as a straight line (almost, because the vaccination
takes place over a finite period of time rather than instantaneously),
indicating a constant elasticity of coverage (at 158 weeks, they were
0.048 in Artibonite and 0.092 in Ouest; see Figures \ref{Fig11} and \ref{Fig12} respectively.) and that the potential benefit to
each person remains constant (herd effects were minimal). The optimal amount
to have vaccinated at 80 weeks would have been three times what was done in
Artibonite and ten-fold greater in Ouest. However, the costs of vaccinating
every person at risk (including those that are or were asymptomatic) is
certainly not a linear function, and a cost-benefit analysis would be necessary
to determine if, and at what point, the money and efforts would be better
expended in other control measures.

\subsubsection{Changing the Timing of Vaccination}

\begin{figure}[H]
\centering\includegraphics[width=6.5in,height=5.1in]{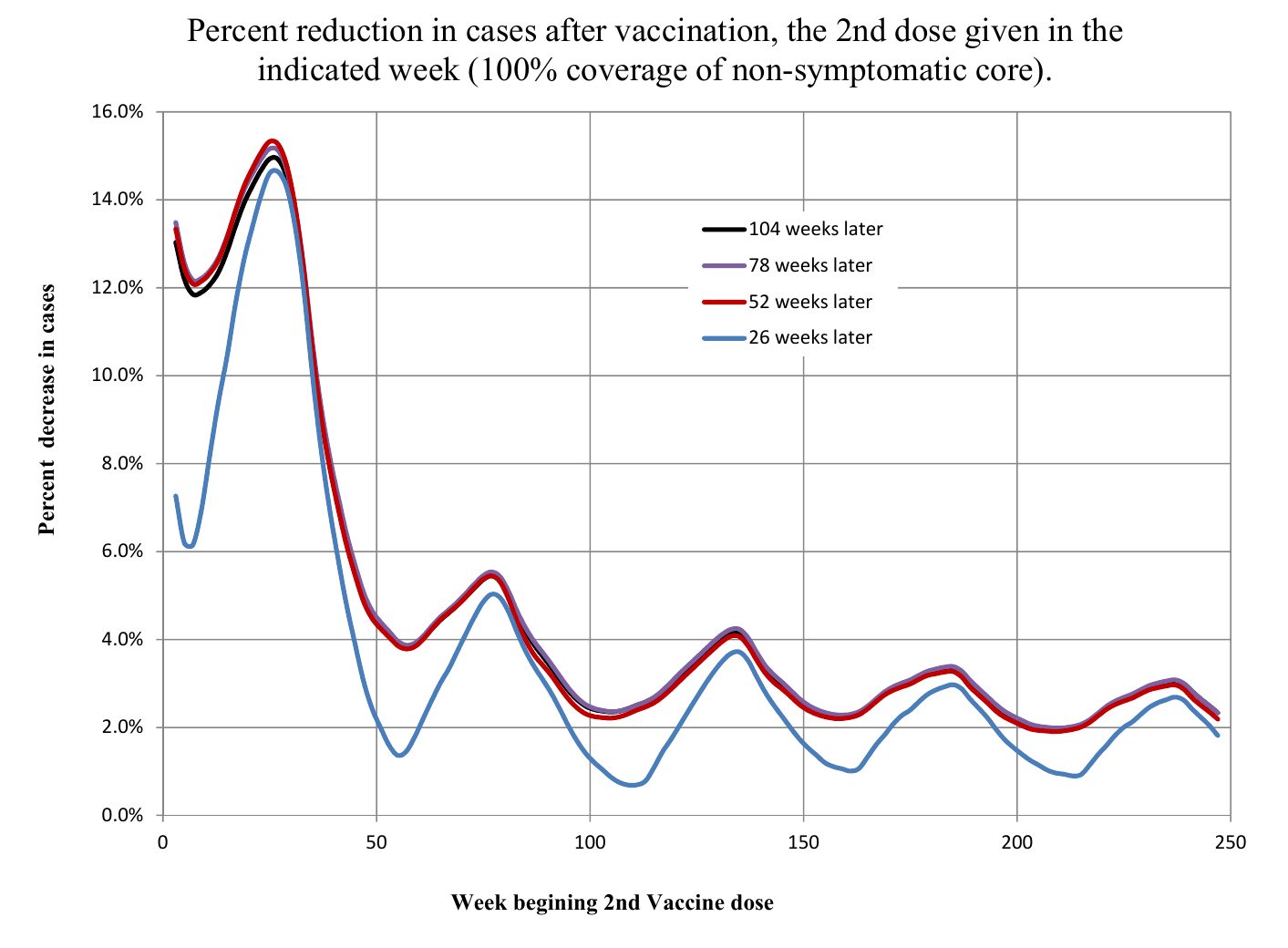}\caption{{ {Artibonite.
The percent decrease in the total number of cases 26, 52, 78 and 104 weeks
after the beginning of the second round of vaccination indicated on the abscissa.
The onset of the immune response was assumed to begin a week later. We assume
$70\%$ efficacy of the vaccine.}}}%
\label{Fig15}%
\end{figure}

\begin{figure}[H]
\centering\includegraphics[width=7in,height=5.5in]{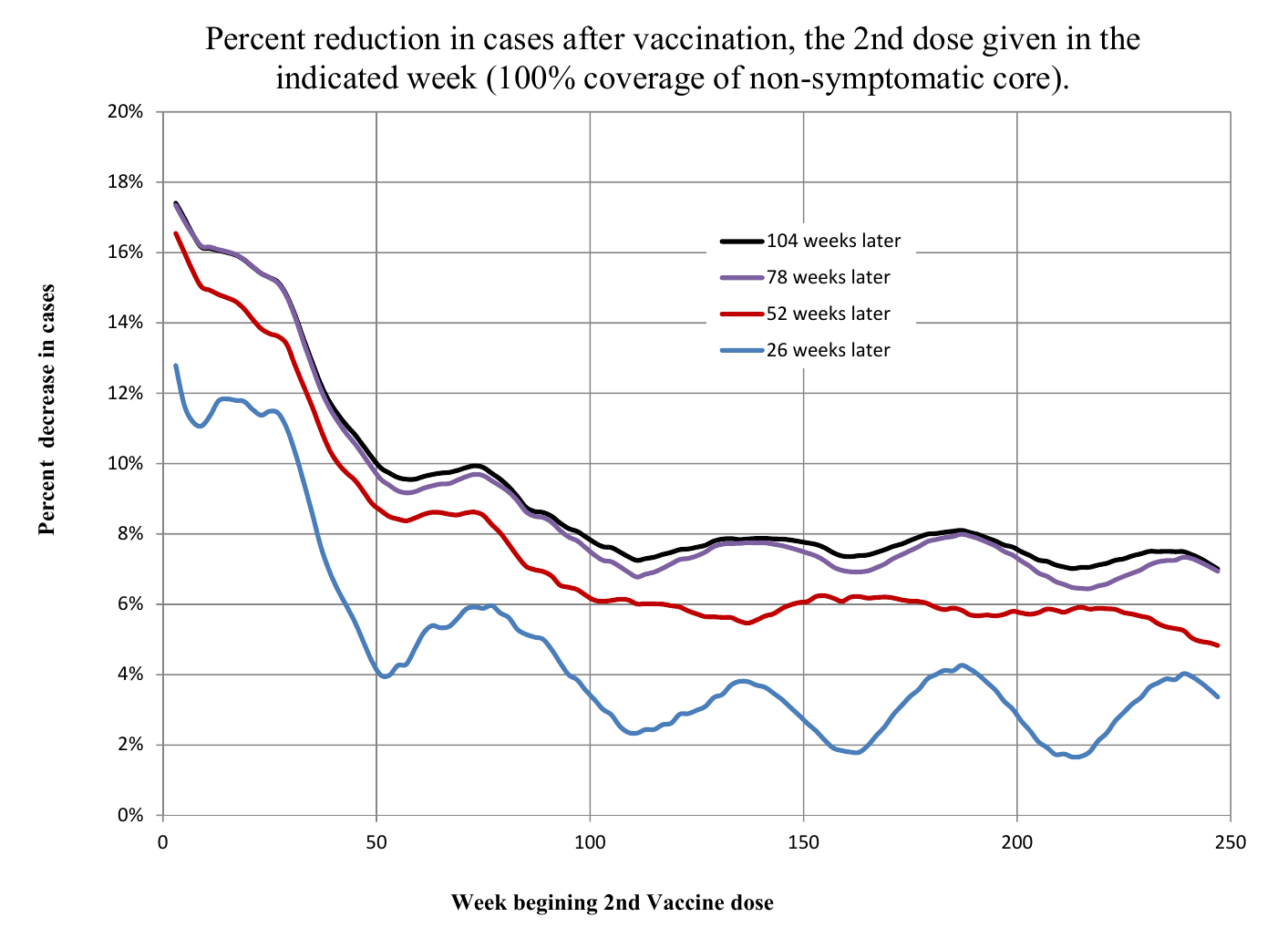}\caption{{ {Ouest.
The percent decrease in the total number of cases 26, 52, 78 and 104 weeks
after the beginning of the second round of vaccination indicated on the abscissa.
The onset of the immune response was assumed to begin a week later. We assume
$70\%$ efficacy of the vaccine.}}}%
\label{Fig16}%
\end{figure}

To investigate the best timing of vaccination, we ran two scenarios: the first,
vaccination\linebreak numbers at $100\%$ coverage of non-symptomatic individuals; and
the second, vaccination near the numbers actually vaccinated. We graph percent
reduction in cases over scenarios without vaccination. \ In Figures
\ref{Fig15} and \ref{Fig16}, for example, the curves represent the percent
reduction in cases as a function of the week in which the second dose was
beginning to be administered. \ There are four different curves on each graph
representing the percent reductions at given intervals (%
${\frac12}$%
, 1, 1%
${\frac12}$%
 and 2 years) after the second dose of vaccination was begun. We simulated
the timing of vaccination so that the beginning of the second round increased
in weekly increments from the third week to the
247th week.


For the first scenario, we simulated the vaccination of the number of
non-symptomatic individuals (susceptible, asymptomatic infected and recovered
asymptomatic infected) in the core in Artibonite and Ouest. There are two
cases for optimizing the timing of vaccination. The first case is to
vaccinate as soon as possible after the epidemic begins. Some of the largest
percent reductions in the number of cases occur when the vaccine second dose is
given between the third and 12th weeks
(between 7 November 2010, and 9 January 2011). The second case is to vaccinate in
early to mid-spring. Vaccination between the 23{rd} and
25{th} weeks (27 March 2011, to 10 April 2011) is both early and
seasonal and has the strongest response. As the weeks roll on, the effects of
vaccine diminish with subsequent local maxima in occurrence between Weeks
73--77 (11 March 2012, to 8 April 2012), 133--137 (5 May 2013, to 2 June 2013),
185--187 (4 May 2014, to 18 May 2014) and 237--239 (3 May 2015, to 17 May 2015);
see Figures \ref{Fig15} and \ref{Fig16}. There is a 46\% drop between the
first spring peak reduction in cases (2011) and the second spring peak
reduction in cases (2012) in Artibonite and a 35\% reduction in Ouest. After
about the 35{th} week, percent case reduction subsequently
drops about 14\% per year in Artibonite and 10\% per year in Ouest.

\begin{figure}[H]
\centering\includegraphics[width=7in,height=6in]{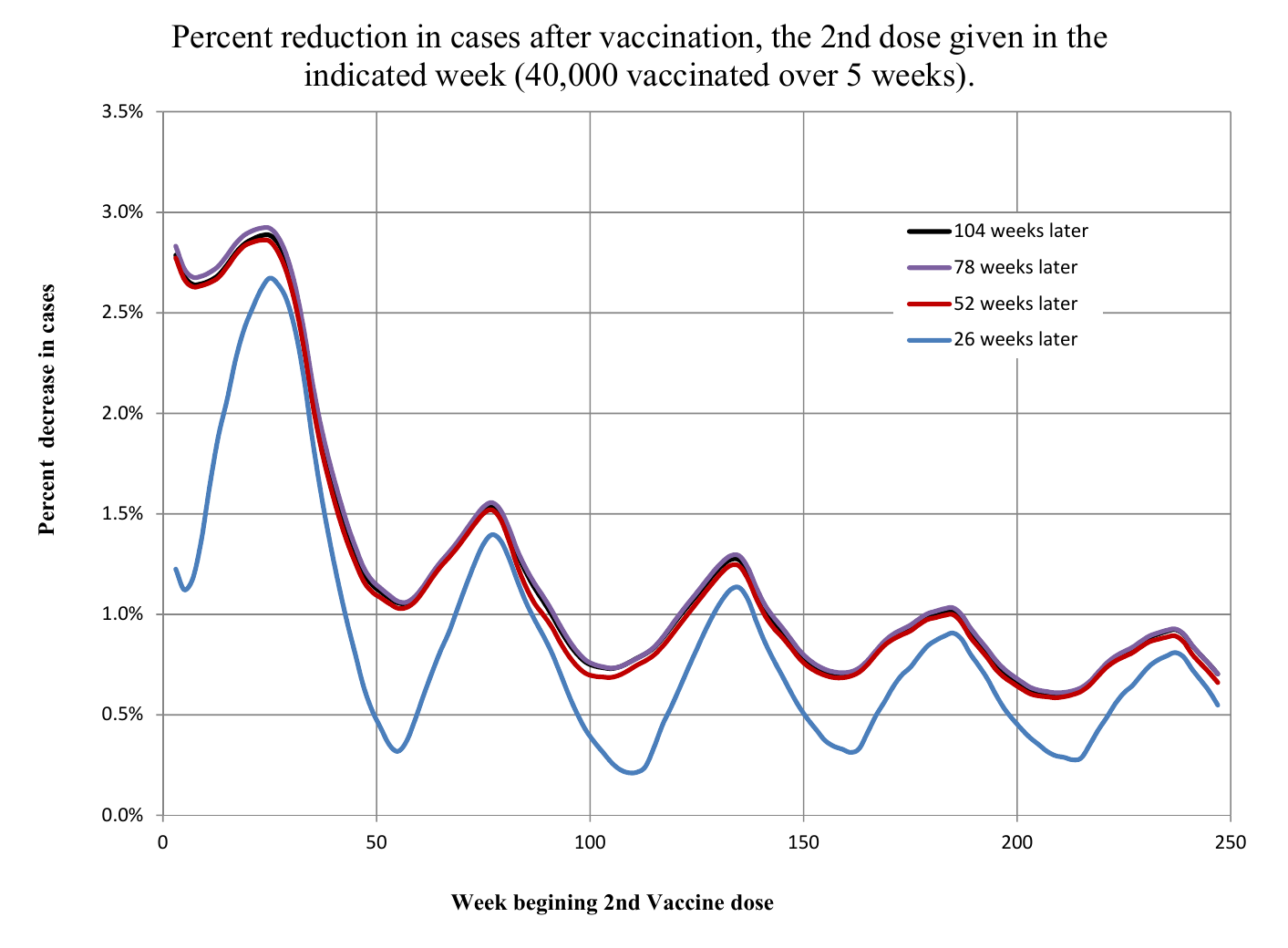}\caption{{ {Artibonite.
The percent decrease in the total number of cases 26, 52, 78 and 104 weeks
after the beginning the second round of vaccination indicated on the abscissa.
The onset of the immune response was assumed to begin a week later. We assume
$70\%$ efficacy of the vaccine.}}}%
\label{Fig17}%
\end{figure}
\begin{figure}[H]
\centering\includegraphics[width=7in,height=6in]{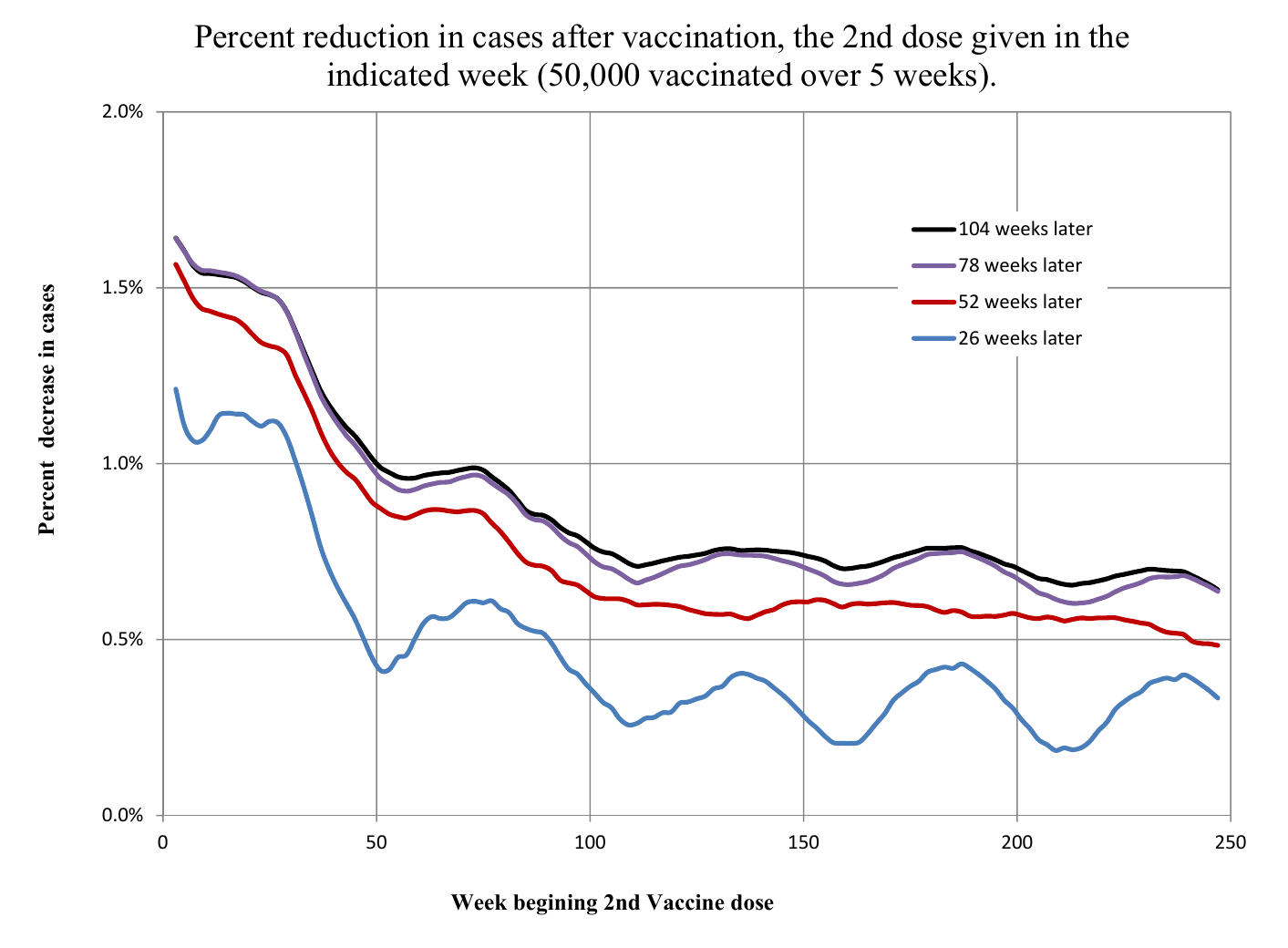}\caption{{ {Ouest.
The percent decrease in the total number of cases 26, 52, 78 and 104 weeks
after the beginning the second round of vaccination indicated on the abscissa.
The onset of the immune response was assumed to begin a week later. We assume
$70\%$ efficacy of the vaccine.}}}%
\label{Fig18}%
\end{figure}

The second scenario (vaccination of 40 thousand in Artibonite and 50 thousand
in Ouest) results in similar cases and timing of optimal vaccine
administration. See Figures \ref{Fig17} and \ref{Fig18}. The amounts of
reduction are not only much lower, but the relative effects are reversed.
\ That is, in the first scenario (near optimal numbers), the percent reduction
in cases after two years in Artibonite start at 12\%--13\% when vaccination begins
at the outset and decline to 3\%--4\% \ when vaccination begins after 4%
${\frac34}$
years. In Ouest, the percent reduction in cases after two years start at 22\%--23\%
when vaccination begins at the outset and decline to 9\%--11\% \ when vaccination
begins after 4%
${\frac34}$
years.

In the second scenario (near actual numbers), the percent reduction in cases
after two years in Artibonite starts at 2.6\%--2.9\% when vaccination begins at the
outset and declines to 0.6\%--0.9\% \ when vaccination begins after 4%
${\frac34}$
years. In Ouest, the percent reduction in cases after two years start at
1.6\%--1.7\% when vaccination begins at the outset and decline to 0.5\%--0.7\%
\ when vaccination begins after 4%
${\frac34}$
years.

The actual timing of the vaccination program was approximately Weeks 81 to 83
for adults and 86 to 87 for the second dose. This was about eight to ten weeks
past the seasonal optimum, which is not quite halfway between the local crest
and trough. Ideally, the vaccination should have been done within a half a
year of the start of the epidemic.

\renewcommand{\thefootnote}{\fnsymbol{footnote}}

\newpage

\section{Discussion}

Modeling the dynamics of cholera in Haiti has been hampered by the lack of
easily accessible, detailed historical meteorological data. We use NASA
satellite data to address this problem. This study shows that with
environmental data of sufficient detail and quality, projections of disease
progression can be made with sufficient lead time to prepare for outbreaks.
The lag times of over five weeks means that if even rudimentary, but reliable,
meteorological and coastal records are kept, preparations and resources can be
more focused. The gathering of basic weather information is simple and
inexpensive and should be made a standard procedure when any agency takes part
in interventions, particularly when the environmental component of the
epidemiology is so well established.

In addition, we explored the hypothesis that, at least in the Ouest region,
tidal influences play a significant role in the dynamics of the disease. It
appeared that tidal range rather than the height of the tide itself had the
strongest influence. Some connection to tidal influences should be expected
where large populations are in close contact with bays and estuaries and
humans are consuming local \linebreak seafood \cite{Huq, deMagny}. It is not surprising
that there was no effect of tidal range found in Artibonite, since the tide
model was for off the coast of Port-au-Prince. Again, the lack of readily
available detailed historical tide records or even a model for various regions
along the coast hinders a thorough investigation of possible factors in the
disease dynamics.

Studies from Africa \cite{Reyburn} found longer time lags (eight to $10$ weeks) and in Bangladesh \cite{deMagny} shorter ones (four weeks). The delays we found in the effects of precipitation on the infection rates
were for Artibonite between $3.4$ and $8.4$ weeks and for Ouest between $5.1$
and $7.4$ weeks, similar in scale to those studies. However, very short delays (four to seven days) have been reported in a
recent study of rainfall forcing in the Haiti cholera epidemic
\cite{Eisenberg}. \ This study limited the time spans investigated to
under 20 days and only during the first year of the epidemic. The authors of that paper also note the discrepancy and suggest that it may reflect the differences between endemic \textit{versus} epidemic situations. An endemic situation would be dominated by rainfall driving transmission through a series of steps, such as washing nutrients leading to plankton blooms, whereas in an epidemic situation, rainfall can bring the population in direct and immediate contact with raw sewage \cite{Eisenberg}. It would be interesting to investigate whether the delays have become longer as the epidemic has proceeded. \ It is certainly plausible, since the force of infection was so much higher at the beginning of the epidemic.

Over the course of the epidemic, the incidence has been tapering off. There has
been steady and continued effort to improve hygiene and living conditions;
however, the areas where the greatest strides are made are those where people
leave the camps to return to normal living conditions and employment. The
declining numbers of those at-risk in the overall population belie the fact
that many local populations are still without basic hygienic facilities. This
was reflected in the model by inclusion of the core group, which comprised
about 9\% of the population in Artibonite and 14\% in Ouest. The initial
values for these percentages were the $u_{0}$ parameters. (The core's
percentages varied very little during the course of the simulations. In both
departments, the difference between maximum and minimum was only about
$0.01\%.$) These values are somewhat higher than the approximate 5\% of the
population that OCHA (Office for the Coordination of Humanitarian Affairs - United Nations)
 reported \linebreak in 2012 still displaced from the 2010 earthquake
\cite{OCHA}. \ Of course, the OCHA number is for the entire country of Haiti,
not just Ouest and Artibonite.

On top of predicting when and how many cholera cases will increase with
Haiti's weather patterns and tides, any modeling to predict the effectiveness
of interventions (such as vaccination) should consider these patterns.
Considering that cholera may be maintained in the environment outside of the
human chain of infection is essential to planing effective prophylaxes and interventions.

Using these models, we were able to do a basic assessment of the relative
effectiveness of the recent vaccination program in Artibonite and Ouest. The
discrepancy between the apparent effectiveness of vaccination in the two
regions is perhaps not that puzzling when one considers the number vaccinated
relative to the size of the at-risk population. In Artibonite, about 41,000
people and in Ouest about \linebreak 47,000 people received both doses of the
vaccine. However, our model suggests that in Artibonite, the at-risk population
by May 6 was about 65,600, 59,400 of which were in the core
population. In Ouest, the at-risk population was still over 379,000 with
295,000 in the core, almost five times the number in Artibonite. In
addition, there were approximately 578 asymptomatic core cases in Artibonite
and 3439 in Ouest, all of whom would have been eligible to receive the
vaccine. Thus, in Artibonite, $69\%$ (40,799) of the most at-risk population
apparently received the vaccine, while in Ouest, only $16\%$ (46,974) did.
Further, in both regions, 100 people receiving vaccination does not mean 100
people protected. The vaccine was about $68\%$ effective in Artibonite and
about $70\%$ effective in Ouest. Therefore, as a rough calculation, we might expect
$0.68\times40,799\approx27,539$ in Artibonite and $0.70\times46,974\approx
32,882$ people in Ouest protected directly.

Yet, of the 40,000 vaccinated in the Artibonite experiments, there is at most 9\%
 (3529/40,000) of those vaccinated not resulting in
cases two years hence, leading to a slightly less than 3\% reduction in
the overall number of cases (Figure \ref{Fig17}). This was vaccination at the
earliest optimal date. Waiting until the week when actual vaccination occurred,
there is about 5\% (1919/40,000) of those vaccinated not
resulting in cases leading to a 1.4\% reduction in the overall number of cases.
Thus, for 40,000 vaccinated, 30,000 are protected, and over the
following two years, the total number of cases reduced due to this protection is
about 2000. In the Ouest experiment, the 50,000 vaccinated there are about
8\% (4037/50,000) of those vaccinated not resulting in
cases leading to a 1.6\% reduction in the overall number of cases (Figure
\ref{Fig18}). Again, this was vaccination at the earliest optimal date. Waiting
until when actual vaccination occurred, there is about 5.2\% (2615/50,000) of those vaccinated not resulting in cases leading
to about a 1\% reduction in the overall number of cases.

\section{Conclusions}

The complex environmental patterns incorporated in epidemic models allow us to remove a major confounding source of variability and highlight the effectiveness of intervention efforts by identifying deviations from the unaltered flow of events. Contaminated waters persist even when the numbers of infected decline. The resulting dynamics imply that while vaccination can reduce infections and suffering over the span of imparted immunity, without large scale, early, and repeated vaccination, elimination of endemic cholera will only come from substantial improvements in living conditions and public hygiene in impoverished communities.


\section*{\noindent Acknowledgments}

The authors would like to thank Scott Braun and George Huffman at NASA, Goddard Space Flight Center, for help accessing precipitation data.

The authors are also grateful to Scott Dowell and Jordan Tappero at CDC, Louise Ivers at PIH and Jean Pape at GHESKIO for information about the cholera vaccination programs in Haiti.

This project was partially supported by a CCFF grant of the Columbian College of Arts and Sciences, of the George Washington University. Svetlana Roudenko was also partially supported by NSF CAREER Grant $\#1151618$ and Stephen Tennenbaum by NSF's Career-life Balance Initiative $\#1250639$.

\section*{\noindent Author Contributions}
\vspace{12pt}

Caroline Freitag and Svetlana Roudenko conceived the project and provided the conceptual framework and early modeling efforts. Stephen Tennenbaum and Svetlana Roudenko further developed and refined the model and did the statistical analysis. Stephen Tennenbaum and Caroline Freitag provided background research, data acquisition, and wrote the manuscript with assistance of \linebreak Svetlana Roudenko.

\section*{\noindent Conflicts of Interest}
\vspace{12pt}

The authors declare no conflict of interest.



\end{document}